\documentclass[a4paper]{ar-1col}

\usepackage{natbib}
\usepackage{lineno}
\usepackage[version=4]{mhchem}
\usepackage{amsmath,amssymb}
\usepackage{textcomp}
\usepackage{subfig}
\usepackage{url}

\jname{Xxxx. Xxx. Xxx. Xxx.}
\jvol{AA}
\jyear{2021}
\doi{10.1146/((please add article doi))}

\begin{document}

\markboth{Wordsworth \& Kreidberg}{Atmospheres of Rocky Exoplanets}

\title{Atmospheres of Rocky Exoplanets}

\author{Robin Wordsworth$^{1,2}$ and Laura Kreidberg$^3$
\affil{$^1$School of Engineering and Applied Sciences, Harvard University, Cambridge, USA, 02138; email: rwordsworth@seas.harvard.edu}
\affil{$^2$Department of Earth and Planetary Sciences, Harvard University, Cambridge, USA, 02138}
\affil{$^3$Max Planck Institute for Astronomy, K\"onigstuhl 17, Heidelberg, Germany, 69117}}

\begin{abstract}
Rocky planets are common around other stars, but their atmospheric properties remain largely unconstrained. 
Thanks to a wealth of recent planet discoveries and upcoming advances in observing capability, we are poised to characterize the atmospheres of dozens of rocky exoplanets in this decade. Theoretical understanding of rocky exoplanet atmospheres has advanced considerably in the last few years, yielding testable predictions of their evolution, chemistry, dynamics and even possible biosignatures. Here we review key progress in this field to date and discuss future objectives. Our major conclusions are as follows:
\newline

\begin{minipage}{4.0in}
\begin{itemize}
\item Many rocky planets may form with initial \ce{H2-He} envelopes that are later lost to space, likely due to a combination of stellar UV/X-ray irradiation and internal heating.
\item After the early stages of evolution, a wide diversity of atmospheric compositions is expected, due to variations in host star flux, atmospheric escape rates, interior exchange and other factors. 
\item Observations have ruled out the presence of hydrogen-dominated atmospheres on several nearby rocky exoplanets, and the presence of any thick atmosphere on one target. More detailed atmospheric characterization of these planets and others will become possible in the near future.
\item Exoplanet biosphere searches are an exciting future goal. However, reliable detections for a representative sample of planets will require further advances in observing capability and improvements in our understanding of abiotic planetary processes.
\end{itemize}
\end{minipage}

\end{abstract}

\maketitle

\tableofcontents

\newpage

\begin{extract}\emph{The Spirits answered, That there were more numerous Worlds than the Stars which appeared...}\\ Margaret Cavendish, The Blazing World, 1688
\end{extract}

\section{INTRODUCTION}\label{sec:intro}

While speculation on the nature of planets around other stars goes back for centuries, exoplanet science is a young discipline, having begun in earnest only in 1995 with the discovery of the hot Jupiter 51 Pegasi b \citep{mayor1995jupiter}. The 26 years since this discovery have seen rapid improvements in instrumentation and observing techniques. These improvements have in turn led to major advances in our understanding of planet formation, the abundance of different planet types around different stars, and the atmospheric compositions of gas-rich planets very different from those in the solar system. Thanks to a combination of ground- and space-based observations, we now also know of hundreds of planets that have around the same radius as Earth \citep{batalha2013kepler}. Most of these planets are distant from the solar system, but a handful of them are close enough to allow more detailed characterization. Some of them may host life.

\begin{marginnote}
\entry{Hot Jupiter}{A gas giant planet that orbits close to its host star and hence has an extremely hot atmosphere.}
\end{marginnote}

Since the first exoplanet discovery, it has been clear that technological advances would eventually allow us to study the atmospheres of these low-mass planets. In anticipation, increasing theoretical attention is now devoted to predicting the composition, circulation, chemistry and climates of rocky exoplanets. In the last few years alone, the very first observational constraints on rocky exoplanet atmospheres have also begun to emerge, and major new steps forward are expected in the near future. 

The motivation for studying rocky planets around other stars is immense. Despite everything we have learned from Earth, Mars and Venus, it is not possible to create a general theory of atmospheres by studying a handful of objects, each of which is very different from the others. Earth, in particular, is unique. No other solar system planets have surface liquid water or, to the best of our knowledge, life. Why is this the case? What determines the distribution of water and other volatiles on rocky planets generally? When do rocky planets retain atmospheres, and why? How does a planet's chemistry affect its evolution, including its chances of developing a detectable biosphere? All these questions were speculative until recently, but in the next few years we will be able to tackle them directly.

Observations of rocky planet atmospheres pose a major technical challenge, because they are dwarfed in size and brightness by their host stars. The most immediately accessible systems are rocky planets transiting small M-dwarf hosts; for these systems, the expected atmospheric signal is $\sim10$ parts per million relative to the host star brightness. A handful of these planets have already been studied with current telescopes, and more observations in greater detail will be enabled by near-future observing facilities such as the James Webb Space Telescope and 30-meter class ground-based telescopes \citep{snellen2015, morley2017}. Direct imaging of reflected light spectra of Earth analogues is also a high priority for the exoplanet community, but these signals are even smaller --- the contrast between the Earth and the Sun is $10^{-10}$ in optical light --- and will require a next generation space telescope \citep{feng2018, charbonneau2018NAS}. 

\begin{marginnote}
\entry{M-dwarf}{A star that is relatively small and cool compared to the Sun. M-dwarfs are the most common type of star in the Milky Way galaxy.}
\end{marginnote}

This review summarizes major results in the study of rocky exoplanets to date, with a focus on the composition and evolution of their atmospheres. By necessity, our discussion is weighted to theory, because only a few observational constraints on rocky planet atmospheres currently exist.  However, in this decade our view of exoplanet atmospheres will undergo a revolution, thanks to the arrival of James Webb and the new ground-based telescopes. By summarizing the current state of knowledge now, our aim is for this review to serve both as an aid for future investigation, and a summary of key predictions of the field ahead of this new wave of observations.

The rest of this review is structured as follows. The following three subsections cover some background and fundamentals. In Section~\ref{sec:atm_escape} we then review the processes that drive atmospheric formation and loss to space. In Section~\ref{sec:chemistry}, we discuss atmospheric chemistry, including the important topic of redox evolution. We discuss atmospheric exchange with the surface and interior in Section~\ref{sec:interiors}, and atmospheric dynamics in Section~\ref{sec:atm_dyn}. Section~\ref{sec:observations} reviews all the observational constraints on rocky exoplanet atmospheres obtained to date. Finally, in Section~\ref{sec:future}, future prospects for the field are discussed, including the exciting question of how we can characterize habitability and search for life on exoplanets.

\subsection{What is a rocky planet?}\label{sec:whatisrocky}

In the solar system, the distinction between rocky planets and gas/ice giants is obvious: Mercury, Venus, Earth and Mars are rocky, while Jupiter, Saturn, Uranus and Neptune are giants. However, a major contribution of exoplanets to planetary science so far has been to show us that this neat solar system distinction does not hold true in general. In fact, planets intermediate in size between Earth and Neptune are an abundant outcome of planet formation \citep{howard2012occurrence}.  Categorizing these worlds is a challenge. Are they scaled-down versions of Neptune, water-rich worlds, or relatively dry rocky planets surrounded by puffy hydrogen envelopes?

Mass and radius measurements can provide useful guidance about the bulk composition of a planet --- most notably, the addition of hydrogen rapidly increases the radius for given mass \citep{valencia2010corot}. This information has been widely used in the exoplanet literature to separate rocky and non-rocky planets, in the absence of any observational constraints on their atmospheric properties \citep[e.g.][]{rogers2015rocky}. Following this approach, we consider a planet rocky if its mass and radius are consistent with that of a bare rock, within typical observational uncertainties for exoplanets (10\% precision on the mass, 5\% precision on the radius). The upper bound on the radius of a rocky planet is then set by the mass-radius relation for a pure silicate (\ce{MgSiO3}) composition, and the lower bound is set by a pure iron composition. Most rocky bodies are expected to be composed of iron cores and silicate mantles and therefore lie between these two extremes. 

This definition includes all the terrestrial planets of the solar system, as even thick atmospheres like that of Venus contribute just $\sim1$\% to the planetary radius at optical wavelengths. The definition also allows for a wide diversity of possible compositions, including ocean worlds that are entirely covered by a water layer, or planets with a small hydrogen atmosphere \citep[$\lesssim0.1$\% by mass;][]{lopez2014radius}. It does not include planets with thicker hydrogen-rich envelopes or tens of percent \ce{H2O} by mass, which we refer to in this article as sub-Neptunes. As we discuss in Section~\ref{sec:atm_escape}, some small hydrogen-rich exoplanets may share a common origin with rocky planets \citep[see also][ for a recent review]{bean2021subneptune}. 

\begin{figure}[hb]
    \centering
    \includegraphics[width=5in]{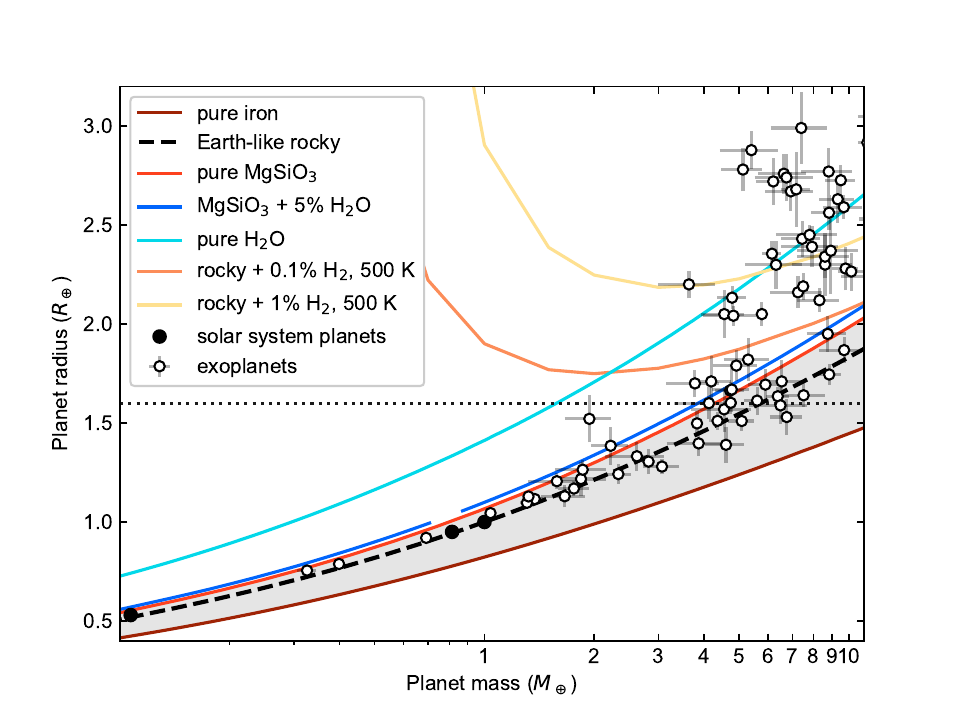}
    \caption{Mass and radius measurements for exoplanets and the solar system planets, compared to model compositions. For clarity we show only planets with 5$\sigma$ mass measurements. The gray shaded region indicates the part of parameter space we consider to be rocky, where the mass and radius are consistent with a composition dominated by iron and/or silicate (here, defined as \ce{MgSiO3}). The horizontal dotted line marks the threshold suggested by \cite{rogers2015rocky}, above which the majority of planets retain a hydrogen envelope. The mass and radius measurements are from the NASA Exoplanet Archive (\protect\url{https://exoplanetarchive.ipac.caltech.edu/}), and model compositions are from \cite{zeng2019growth}.}
    \label{fig:mass_radius}
\end{figure}

Figure \ref{fig:mass_radius} shows the mass-radius relation for exoplanets and the solar system. Planets below roughly 1.6 Earth radii are consistent with a rocky composition \citep{weiss2014, rogers2015rocky, wolfgang2015}. At larger radii, there is much more scatter in the mass-radius relation, with most (but not all) planets requiring a significant mass fraction of volatiles. Exoplanets in the 1.5-2.0~$R_\oplus$ radius range are less abundant than those of lower or higher radius, most likely because of the rapid increase in radius that occurs once a hydrogen envelope is present. This phenomenon has come to be known as the radius gap \citep{fulton2017gap, vaneylen2018gap}. 

\subsection{Key techniques for observing exoplanet atmospheres}

\begin{figure}[ht]
    \centering
    \includegraphics[trim={0 22cm 0 0},clip,scale=0.4]{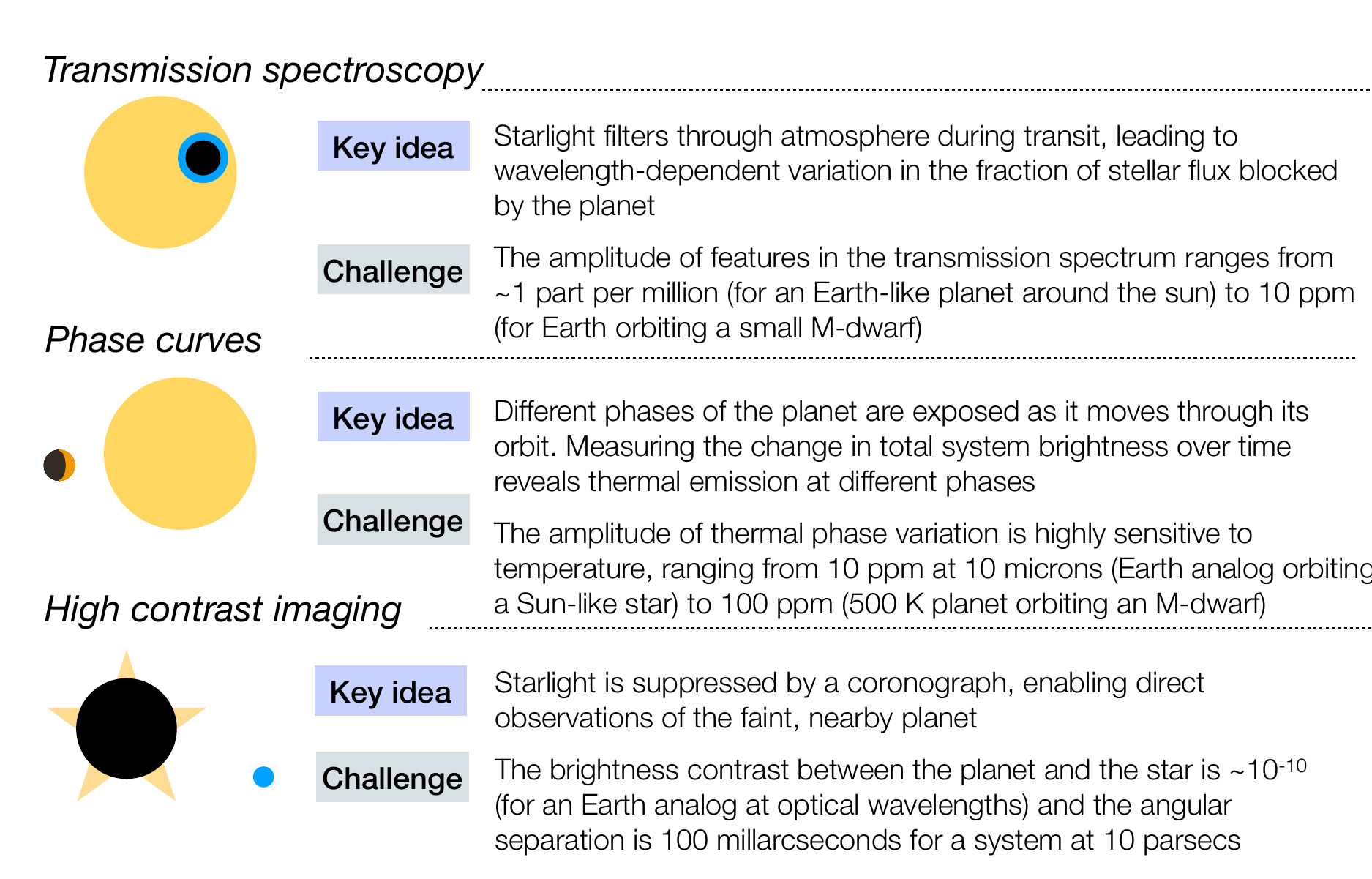}
    \caption{Summary of key techniques for exoplanet atmosphere characterization.}
    \label{fig:techniques}
\end{figure}

There are several general approaches used to measure the tiny signals from rocky planet atmospheres (summarized in the cartoon in Figure\,\ref{fig:techniques}). One is time-based: one can monitor the combined flux from the star and planet over time, to search for variation in brightness as the planet's viewing geometry changes. The most common time-based method is transmission spectroscopy, which consists of a series of spectra measured over the course of a planet's transit. During the transit, a small fraction of the stellar light is transmitted through the planet atmosphere, where it can be absorbed or scattered.   In addition to transit spectroscopy, the planet can also be observed at other times during its orbit. During secondary eclipse, all the thermal emission and reflected light from the planet are blocked by the star, and the planet signal can be inferred from the missing light. The planet can also be observed over its entire orbital period (a so-called phase curve), where different phases of the planet are observable over time. These measurements can be made at low spectral resolution, to reveal broad molecular absorption bands. Alternatively, the atmosphere can be observed at high  resolution ($R\sim100,000$), which spectrally resolves individual absorption lines \citep[e.g.][]{snellen2010}. High-resolution specta are typically blended with light from the host star, but the signal-to-noise can be improved with starlight suppression techniques \citep{lovis2017}.

\begin{marginnote}
\entry{phase curve}{A time series observation of a planet over its entire orbit.}
\end{marginnote}

\begin{marginnote}
\entry{Solar composition gas}{A gas mixture with the same elemental abundance ratios as the Sun.}
\end{marginnote}

These combined light measurements require bright, nearby host stars (generally within 15 parsecs). They are most feasible for planets with small M-dwarf host stars (the smaller the star, the larger the planet signal). They are also best suited to short-period planets, which are statistically more likely to transit. Table\,\ref{tab:best_targets} lists the 10 most observationally accessible rocky planets for combined light measurements (as of 2021). Of particular note is the TRAPPIST-1 system, which has seven rocky planets transiting an ultra-cool dwarf star \citep{gillon2017}. These planets are among the easiest to characterize, opening up detailed comparative study between multiple planets in the same system \citep{morley2017, lustigyaeger2019trappist}.   Extensive surveys from the ground and space \citep[e.g. the Transiting Exoplanet Survey Satellite;][]{ricker2015} have searched a large fraction of the sky with the sensitivity needed to detect such systems, so it is increasingly unlikely that better targets will be found.  The targets listed in Table~\ref{tab:best_targets} may therefore be the most accessible transiting planets we will ever know.

Most observational effort to date has focused on combined light measurements, but to push to rocky planets that have larger host stars and wider orbits (e.g., Earth analogs), the more promising approach is direct imaging: high contrast, high angular resolution observations that suppress the light from the star by many orders of magnitude.  Direct imaging of Earth analogs is beyond that capabilities of current instrumentation, but this is a major goal for future telescopes, as we discuss in Section~\ref{sec:future}.

\begin{table}[]
    \centering
    \begin{tabular}{l|lllllll}
    Planet & $R_p$ & $M_p$ & Period & $T_\mathrm{eq}$  & $R_*$ & Distance & TSM\\
\, & (R$_\oplus$) & (M$_\oplus$) & (Days) & (K) & ($R_\mathrm{sun}$) & (parsec) & \,\\
 \hline\hline
TOI-540 b & 0.9 & 0.7 & 1.2 & 611 & 0.19 & 14.0 & 38.8 \\
Gliese 486 b & 1.3 & 2.8 & 1.5 & 700 & 0.33 & 8.1 & 35.5\\
LHS 3844 b & 1.3 & 2.7 & 0.5 & 804 & 0.19 & 14.9 & 35.2 \\
LTT 1445 A b & 1.3 & 2.9 & 5.4 & 420 & 0.28 & 6.9 & 30.2 \\
GJ 1132 b & 1.1 & 1.7 & 1.6 & 529 & 0.21 & 12.6 & 29.3 \\
GJ 357 b & 1.2 & 1.8 & 3.9 & 524 & 0.34 & 9.4 & 29.2 \\
TRAPPIST-1 b & 1.1 & 1.0 & 1.5 & 399 & 0.12 & 12.1 & 27.6 \\
L 98-59 c & 1.4 & 2.4 & 3.7 & 522 & 0.31 & 10.6 & 26.7 \\
LHS 1140 c & 1.3 & 1.8 & 3.8 & 438 & 0.21 & 15.0 & 25.2 \\
TRAPPIST-1 d & 0.8 & 0.3 & 4.0 & 287 & 0.12 & 12.1 & 24.1 \\
TRAPPIST-1 c & 1.1 & 1.2 & 2.4 & 341 & 0.12 & 12.1 & 23.5 \\
    \end{tabular}\vspace{0.2in}
    \caption{System parameters for the rocky planets most accessible for transmission spectroscopy, based on data from the NASA Exoplanet Archive. The Transit Spectroscopy Metric (TSM) is the normalized signal-to-noise for features expected in the transmission spectrum \citep{kempton2018}. The equilibrium temperature assumes full heat redistribution and zero Bond albedo. The masses of TOI-540b and LHS 3844b were estimated from an Earth-like mass-radius relation. }
    \label{tab:best_targets}
\end{table}

\begin{marginnote}
\entry{volatile}{An element or species found predominately in gaseous form in a given temperature and pressure range}
\end{marginnote}

\begin{marginnote}
\entry{refractory}{An element or species found predominately in solid form in a given temperature and pressure range}
\end{marginnote}

\subsection{Some initial insights from the solar system}\label{sec:insights}

\begin{figure}[ht]
    \centering
    \includegraphics[width=5in]{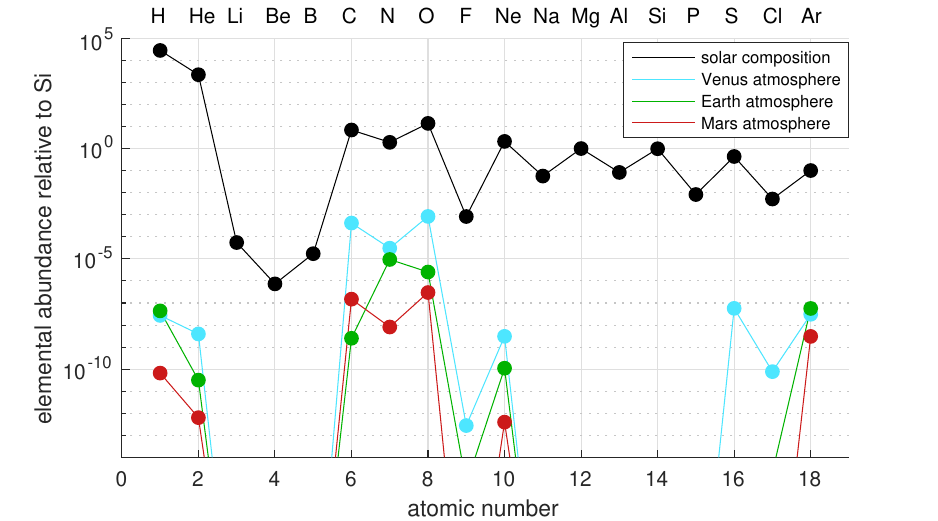}
    \caption{Abundances of low-mass elements in the solar system, relative to silicon (Si), based on values from \cite{lodders2003solar}, \cite{yung1999photochemistry} and references therein. For the atmospheres of Venus, Earth and Mars, abundances are shown relative to mantle silicon. The elemental abundance of the Sun is similar to that of nearby stars that host small planets  \citep{buchhave2015metallicities,bedell2018chemical}. Atmospheric abundances of some minor elements, such as P on Venus \citep{greaves2021phosphine,snellen2020re}, remain poorly constrained.}
    \label{fig:solar_system_abundances}
\end{figure}

While exoplanets are expected to be highly diverse, much has already been learned about rocky planet atmospheres from the solar system. Figure~\ref{fig:solar_system_abundances} shows a plot of the atmospheric abundances of a few key volatile and refractory elements relative to bulk mantle silicon (Si) for Venus, Earth, Mars and solar composition gas. A few major trends are immediately obvious. First, Li, Be, B, Na, Mg, Al and Si are absent from all three terrestrial planet atmospheres. More surprisingly, all three atmospheres are also substantially depleted in volatile elements relative to solar composition: most obviously H and He, but also F, Ne, S, Cl and Ar, and to a lesser extent C, N and O. These underlying elemental abundances lead to the predominance of a handful of key molecules in each atmosphere: \ce{CO2} for Venus, \ce{N2} and \ce{O2} for Earth, and \ce{CO2} for Mars.

The differences seen in Fig.~\ref{fig:solar_system_abundances} are due to a combination of formation, loss  and interior sequestration processes. Most notably, the extreme H and He depletion shows that these gases were either never captured from the nebula, or they escaped to space early on. Other differences are due to sequestration in the interior. Li, Be, B, Na, Mg, Al and Si all behave as highly refractory elements in the 150-800~K temperature range. In addition, due to various chemical processes, all three planets are often thought to have large (albeit poorly constrained) inventories of H, C, N and S in their crusts, mantles and cores \citep[e.g, ][]{desch2020volatile}. A third major effect is condensation on the surface: \ce{H2O} as liquid oceans on Earth, and both \ce{CO2} and \ce{H2O} as ice deposits on Mars. Finally, Earth's atmospheric composition is also strongly affected by the presence of life.  Combined, these processes form the basis for predicting the compositions of rocky exoplanet atmospheres.

\begin{figure}[h!]
    \centering
    \includegraphics[width=5in]{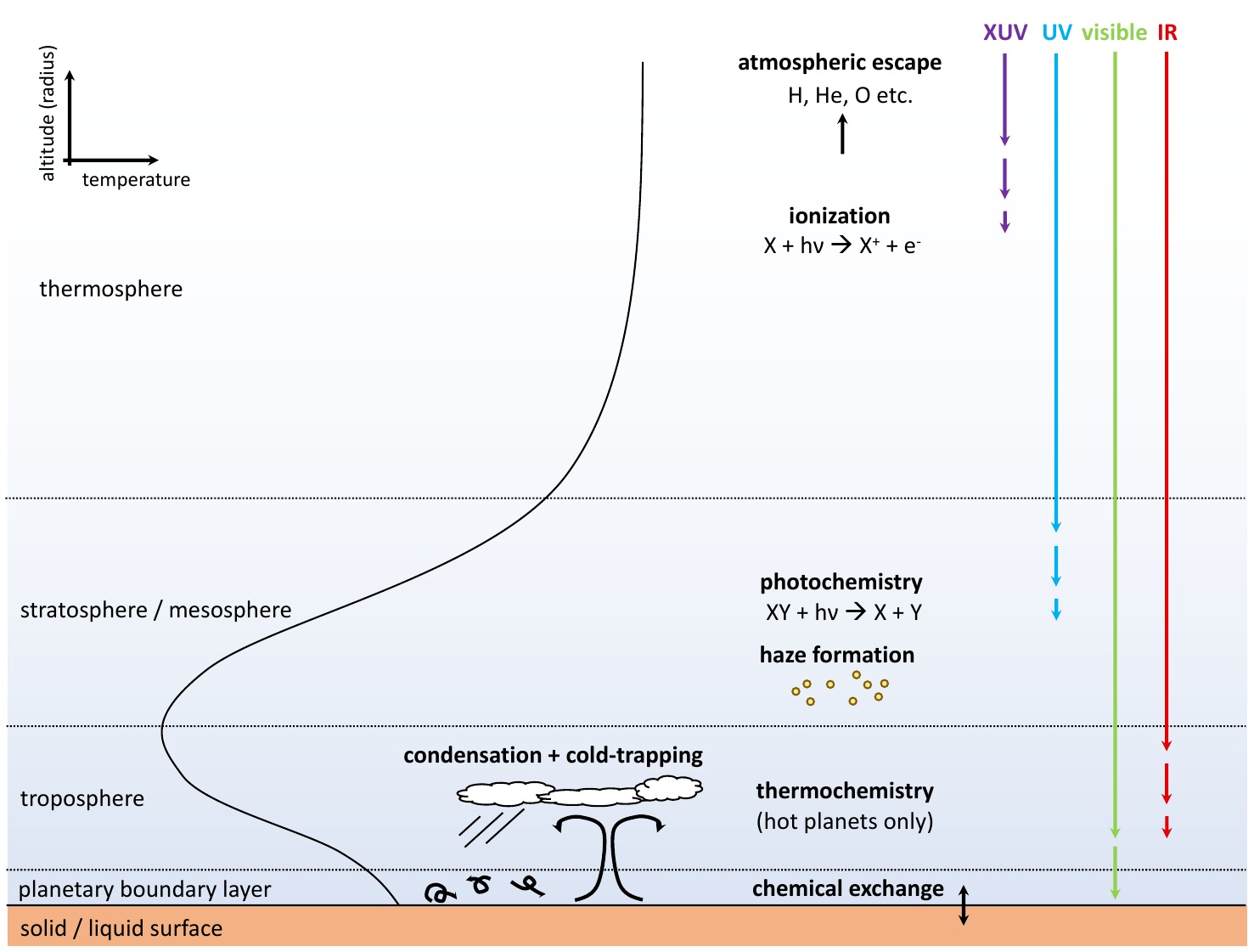}
    \caption{Schematic indicating temperature structure (solid black line), key physical processes and transparency to XUV, UV, visible and IR radiation in the atmosphere of a rocky planet. Note that thermospheric temperatures vary widely depending on stellar flux and atmospheric composition. }
    \label{fig:atm_schematic}
\end{figure}

The thermal structures of solar system atmospheres vary widely, but there are common themes that are expected to hold true for exoplanets. Figure~\ref{fig:atm_schematic} shows a schematic of the temperature structure and key physical processes in a typical rocky planet atmosphere. In most cases, there is a lower region \citep[typically at pressures of $\sim$0.1~bar and greater; ][]{robinson2014common} where convection dominates (the troposphere) and an upper region where convection is inhibited and radiative balance ensues (the stratosphere). Below the troposphere, the highly turbulent planetary boundary layer connects the surface to the lower troposphere. Within the troposphere, temperature declines with altitude due to near-adiabatic behavior of convecting air parcels. In the stratosphere, temperature declines more slowly, and can even increase with altitude due to UV absorption by species such as ozone (Figure~\ref{fig:atm_schematic}). Above the stratosphere lies the thermosphere, where heating by absorption of extreme ultraviolet and X-ray (XUV) stellar photons makes temperatures highly variable, increasing up to thousands of degrees K in some cases. This region is critical to atmospheric escape, as we discuss in Section~\ref{sec:atm_escape}. 

\begin{marginnote}
\entry{XUV}{In the planetary literature, XUV is defined as the combination of X-ray and EUV radiation (i.e., radiation at all wavelengths below 121~nm)}
\end{marginnote}

\begin{marginnote}
\entry{adiabatic}{A process in which there is no heat or mass transfer between a system and its environment}
\end{marginnote}

In the solar system, surface pressures are mainly an outcome of slow evolutionary processes and range from 92~bar (Venus) to $<10^{-4}$~bar (Mercury, Pluto). Surface temperatures, on the other hand, are largely governed by the interaction of the atmosphere with solar radiation on relatively short timescales via the greenhouse effect. The surface temperatures of rocky planets with atmospheres are almost always greater than their equilibrium temperatures, because most atmospheres are more transparent in the visible than in the infrared (IR). As a result, incoming stellar radiation that is absorbed by the surface must be re-emitted as IR radiation higher in the atmosphere, giving rise to greenhouse warming \citep{pierrehumbert2010principles}. 
For thermal radiation in the 100~K to 5000~K range (i.e., 30 to 0.6 microns Planck peak), absorption by typical atmospheric species is due to vibrational-rotational transitions within molecules and collision-induced absorption involving multiple atoms and molecules. Homonuclear, diatomic species such as \ce{H2} and \ce{N2} are weak IR absorbers at low abundance, while species such as \ce{CO2}, \ce{H2O}, \ce{NH3} and \ce{CH4} are strong absorbers, making them effective greenhouse gases. However, even \ce{H2} and \ce{N2} become effective greenhouse gases at pressures above $\sim 0.1$~bar due to collision broadening and collision-induced absorption effects \citep{goldblatt2009nitrogen,wordsworth2013hydrogen}. 

Volatile condensation plays a major role in the composition of all planetary atmospheres.  Over equilibrium temperature ranges from 39~K \citep[Pluto;][]{zhang2017haze} to 2700~K \citep[dayside of 55 Cancri e;][]{demory2016map}, condensables can vary from highly volatile elements like N in the form of \ce{N2}, to relatively refractory elements like \ce{Na}. Only \ce{H2}, \ce{He} and \ce{Ne} can safely be regarded as volatile in all planetary environments. Volatile condensation lets us make broad statements about the atmospheric species permitted for a given equilibrium temperature (Section~\ref{sec:future}). It also plays a more subtle role in atmospheric chemical evolution via the cold-trap effect (Sections~\ref{sec:atm_escape} and ~\ref{sec:chemistry}). 

Closely related to volatile condensation is the runaway greenhouse effect. This process, which may be responsible for the divergent climates of Earth and Venus, occurs when a condensable species is also a strong infrared absorber  \citep[e.g., ][]{ingersoll1969runaway}. It results in a non-linear transition from a mixed volatile-condensate state to a volatile-only state as a planet's equilibrium temperature increases. For \ce{H2O}, the most studied runaway greenhouse species, the runaway transition occurs at around $T_{eq} = 266$~K \citep{goldblatt2013low}. Runaway greenhouse effects are important to determining atmospheric composition, and can also influence a rocky planet's observed radius \citep{turbet2020revised}.

\section{ATMOSPHERIC ACQUISITION AND LOSS TO SPACE}\label{sec:atm_escape}

Formation and loss processes are critical to the evolution of rocky planets over their lifetimes. In this section, we review how atmospheres are acquired during planet formation, and then discuss the processes that drive loss of H and other species to space. 

\subsection{Atmospheric capture and volatile delivery during formation}\label{subsec:atm_delivery}

There are two essential ways that volatile elements can be delivered to planets: as gases in the protoplanetary nebula, or as condensed solids (the latter either in pure form or bound to other elements). 
In solar system science, it is common to refer to the atmospheres that result from these two processes as primary and secondary, respectively. 
Nebular capture occurs when a protoplanet forms before the nebula is dispersed. For gas to remain gravitationally bound to a planet or protoplanet, the temperature of the atmosphere must be low enough that most molecules have kinetic energy $E_K$ lower than their gravitational potential energy $E_P$. The point at which this happens can be defined in terms of a gravitational escape parameter (sometimes called the Jeans parameter) 
\begin{equation}
\lambda \equiv \frac{E_P}{E_K}  = \frac{GMm}{kTR}.\label{eq:lambda}
\end{equation}
Here $G$ is the gravitational constant, $M$ is the planet mass, $m$ is the molecular mass, $k$ is Boltzmann's constant, $T$ is the local gas temperature, and $R$ is the radius of the base of the escaping region (which may be larger than the solid planet radius). When $\lambda$ is large, as is the case for more massive, cooler objects, atmospheric capture from the nebula is favored.

The mass of gas that gets captured is low for small protoplanets with high solid accretion rates, because accretion heats the planet, decreasing $\lambda$. When the solid accretion rate is low, the fraction of nebula captured depends on how much the atmosphere has cooled towards an isothermal profile \citep{hayashi1979earths,lee2015cool,owen2020hydrogen}. At a threshold of roughly 1~$M_E$, protoplanets surrounded by nebula can accumulate \ce{H2}-He envelopes of around 1\% of their mass, greatly increasing their radius and converting them from rocky to gas planets according to our definition in Section~\ref{sec:whatisrocky}.

Geochemical evidence suggests some early nebular capture did in fact occur on the solar system rocky planets, or at least on their building blocks \citep[e.g., ][]{williams2019capture}. However, these thin primitive atmospheres were then rapidly lost to space shortly after the nebula dissipated \citep{lammer2020loss}. For rocky exoplanets, a diverse range of histories is possible. Indeed, many exoplanets in the 1-3~$R_\oplus$ range may represent a population that formed with \ce{H2}-\ce{He}-rich atmospheres but only later bifurcated due to loss to space \citep{vaneylen2018gap}. The fact that the solar system planets apparently formed too slowly to reach the mass threshold where they could accumulate percent levels of \ce{H2} and become sub-Neptunes before the disk dissipated is quite puzzling. Further modeling, exploration of exoplanet population statistics and atmospheric characterization are required to understand this important aspect of solar system evolution.

In the absence of efficient nebular capture, delivery of volatiles that are incorporated into silicate minerals or condensed as ices provides another way to build rocky planet atmospheres. The latter process begins when volatiles condense in the outer portions of protoplanetary disks according to the local gas temperature. The decline in disk temperature both with distance from the protostar and with time allows snow lines for various volatile species to be defined  \citep{oberg2011effects}. Broadly speaking, solid bodies that accrete outside the snow line will be volatile rich, while those that accrete inside will be volatile poor. 

The \ce{H2O} snow line in the solar system occurred at around 2-4~AU \citep{morbidelli2012building}, but snow lines vary with star type. Around M-stars, the long period of elevated luminosity before the star joins the main sequence \citep{baraffe2015new} may lead to lower rates of delivery of \ce{H2O} to close-in rocky planets compared to the solar system \citep{unterborn2018inward}. However, formation inside a snow line does not preclude the presence of that species on a planet. Earth is not completely dry today, either because \ce{H2O} was present in small quantities in its building blocks, or it was delivered from other sources, such as late-accreting planetary embryos \citep{raymond2004making}. In the latter case, the stochastic nature of accretion and variability in planetesimal water content due to radiogenic heating means  that high variance in rocky planet initial water content is expected \citep[e.g., ][]{lichtenberg2019water}. Planets can also migrate inward from their original formation locations, as may have occurred in the TRAPPIST-1 system \citep{unterborn2018inward}.

\subsection{Loss of volatiles to space: Hydrogen}

As soon as the protoplanetary disk disperses, young rocky planets begin to lose their atmospheres to space. Depending on the circumstances, atmospheric loss may slow to negligible rates after a short time, or it may continue for the lifetime of the planet. It can strip away the entire atmosphere, or leave it almost untouched. Understanding this process is hence vital to understanding the range of possible outcomes for rocky planets. 

All loss processes work by overcoming the force of gravitational attraction that keeps an atmosphere bound to its planet. In hydrodynamic escape, the atmosphere flows outward to space continuously, as a fluid. Hydrodynamic escape is in a sense the reverse of nebular capture, and it becomes important once $\lambda$ has values of around 10 or lower. For this to happen, a planet's upper atmosphere must be hot. To get a sense of the numbers, for an Earth-mass planet with extended \ce{H2} atmosphere, at a distance $R=1.3 R_\oplus$, $\lambda = 10$ if $T = 1150$~K. 
Over one thousand Kelvins appears high compared to, say, Earth's equilibrium temperature of 255~K, but such temperatures are quite easily reached in the upper atmospheres of low-mass planets. One major reason for this is XUV radiation \citep{zahnle1982evolution}. Most gases are highly opaque to XUV radiation (Fig.~\ref{fig:atm_schematic}), so it gets absorbed high up, where density is low. In \ce{H2}-dominated atmospheres, the absorbed radiation cannot be easily re-radiated at temperatures below around 10,000~K \citep{murray2009atmospheric}, and conduction to the lower atmosphere is often inefficient \citep{watson1981dynamics}. Hence extremely high temperatures and hence low $\lambda$ values are reached, which in turn leads to hydrodynamic outflow to space. 

In the limit where most XUV energy is used to drive atmospheric loss to space, an energy-limited mass loss flux can be calculated as
\begin{equation}
\phi_E = \frac{\epsilon F_{XUV}}{4 V_{pot}}.
\end{equation}
Here $F_{XUV}$ is the planet's received XUV flux, $V_{pot}=GM/R$ is the gravitational potential magnitude at the base of the escaping region, and $\epsilon$ is an efficiency factor. XUV-driven loss is predicted to be fast on young planets around active stars: for $\epsilon = 0.1$, $F_{XUV} = 10^{-3}$ times the bolometric flux, and other parameters appropriate to early Earth, $\phi_E = 3.8\times10^{-10}$~kg~m$^{-2}$~s$^{-1}$, implying loss of an entire ocean's worth of H ($2\times 7.6\times10^{22}$~mol) in just 25~My.

\begin{marginnote}
\entry{bolometric flux}{Total flux integrated over all wavelengths.}
\end{marginnote}

XUV-driven hydrodynamic escape is regarded as one of the most important escape mechanisms for \ce{H2} from low-mass planets. Indeed, the 1.5-2.0~$R_\oplus$ radius gap \citep{fulton2017gap, vaneylen2018gap} was predicted theoretically a few years before it was observed in models that invoked this escape mechanism \citep{lopez2013role,owen2013kepler}. 
Over the last few years, however, it has become clear that many exoplanets may also lose large amounts of hydrogen due to a related process that has come to be known as core-powered mass loss\footnote{In this context, the core refers to the combined rocky and iron-dominated portions of the planet.} \citep{ikoma2012situ,ginzburg2018core,gupta2019sculpting}. In brief, core-powered mass loss is hydrodynamic escape that is driven by thermal energy from the planet's troposphere and deeper atmospheric regions, rather than high-energy stellar photons absorbed in the thermosphere. 

Core-powered mass loss and XUV-driven escape may both be important escape mechanisms for hydrogen from low-mass planets. However, there are important differences in the timescales on which they will operate. For most host star types, XUV fluxes decline rapidly on a timescale of $\sim 100$~My or less after formation, whereas core-powered mass loss is believed to act on longer timescales of $\sim$1~Gy \citep{lopez2013role,gupta2019sculpting}. Recently, an analysis of the Gaia-Kepler stellar properties catalogue found significant differences in sub-Neptune vs. rocky planet populations between systems with an age cut-off of 1~Gy, suggesting that the role of core-powered mass loss is significant \citep{berger2020gaia}. However, further analysis is required to definitively determine the key loss mechanism responsible for converting sub-Neptunes into rocky planets. 

\begin{marginnote}
\entry{homopause}{Level in the atmosphere at which eddy diffusion equals molecular diffusion. Above the homopause, the atmosphere is not well-mixed and gas species begin to separate according to their masses.}
\end{marginnote}

Hydrogen loss can still occur even after \ce{H2} is no longer the dominant species in a planet's atmosphere. In these circumstances, the bottleneck in loss usually becomes the rate at which hydrogen-bearing species such as \ce{H2O} or \ce{CH4} can diffuse through the homopause in the upper atmosphere, as determined by the formula 
\begin{equation}
\Phi_1 = b x_1 \left(H_{s,2}^{-1} - H_{s,1}^{-1}\right).\label{eq:diff_lim_escape}
\end{equation}
Here $\Phi_1$, $x_1$ and $H_{s,1}$ are the molecular flux, molar concentration and scale height of the hydrogen-bearing minor species, $H_{s,2}$ is the scale height of the background atmosphere, and $b$ is a binary diffusion coefficient \citep{hunten1987mass}.  This regime is called diffusion-limited escape. 

Once hydrogen-bearing species reach the upper atmosphere, they are usually photolyzed rapidly, releasing H that then escapes to space. In some cases, the flux of the hydrogen-bearing species from the surface depends on chemical or even biological factors \citep[e.g., \ce{CH4} on the early Earth; ][]{catling2001biogenic}. However, the most important carrier of H on \ce{H2}-poor planets is usually \ce{H2O}, which is readily cold-trapped in the lowest regions of the atmosphere when a planet's equilibrium temperature is below the runaway greenhouse limit (Section~\ref{sec:insights}). Cold-trapping of \ce{H2O} will protect against H loss, as long as the planet's inventory of non-condensing gases such as \ce{N2} is sufficient \citep{wordsworth2013water}. When cold-trapping fails, continued H loss can dramatically alter a rocky planet's chemistry over time, as we discuss further in Section~\ref{sec:chemistry}.

\subsection{Loss of volatiles to space: Heavier species}\label{sec:atm_escape_C}

Elements heavier than hydrogen are harder to lose to space on terrestrial planets, which is fortunate for us, because it permits the survival of life on Earth. However, there are still many processes that contribute to heavy element loss. The first is conceptually straightforward: these elements can simply be dragged along with the escaping \ce{H2} during hydrodynamic escape. Indeed, in a well-mixed gas undergoing rapid escape, all elements will be lost in equal proportions. However, for moderate escape rates, escaping gases have time to diffusively separate, causing the heavier atoms or molecules to escape less rapidly. Below a certain critical escape flux, the loss of the heavier species is predicted to shut down entirely \citep{hunten1987mass}.

The general equations for diffusion in a mixture of gases \citep{chapman1970mathematical} allow equations for hydrodynamic drag to be derived from first principles \citep{zahnle1990mass,wordsworth2018redox}. This conveniently allows the critical mass flux $\phi_c$ required to initiate drag of a heavy species along with a light species to be derived directly. It can be shown that this mass flux is simply $\phi_c = m_1 \Phi_1$, with $\Phi_1$ defined in \eqref{eq:diff_lim_escape}. Above the critical mass flux, the escape rate of the heavier species 2 is approximately
\begin{eqnarray}
\Phi_2 \approx \left\{
  \begin{array}{lr}
   0 & :  \phi  < \phi_c  \\
    \left[x_2\phi +  x_1 x_2(m_1-m_2)  b/H_{s,1} \right]/ {\overline m}   & :  \phi  \ge \phi_c
  \end{array} \right.
      \label{eq:num_flux}
\end{eqnarray}where  $\overline m = m_1 x_1 + m_2 x_2$ is the mean molecular/atomic mass of the flow.  Equation \ref{eq:num_flux} is exact in the limit of subsonic, isothermal flow, and it is reasonably accurate in more general situations \citep{zahnle1990mass}. It shows that the mass difference between the two species and the scale height of the lightest species are both important to determining the critical mass flux. A complementary expression for the escape rate of the lighter species can also be derived that reduces to \eqref{eq:diff_lim_escape} when $\phi<\phi_c$.

Hydrodynamic drag matters for exoplanets because it influences the composition of a rocky planet's atmosphere after its primordial \ce{H2} envelope has been lost to space \citep[e.g., ][]{malsky2020coupled}. Because the degree of heavy element fractionation depends strongly on the rate of \ce{H2} loss through time, atmospheric characterization may provide an additional way to distinguish between various loss mechanisms. Even direct measurements of isotope fractionation may be feasible in the future: it has been suggested that high-resolution ground-based or JWST transit observations could be used to detect highly deuterium-enriched water on rocky exoplanets such as Proxima Cen b or TRAPPIST-1b \citep[][see also Section~\ref{sec:future}]{molliere2019detecting,lincowski2019observing}.

Once most \ce{H2} is lost to space, atmospheres become far more resistant to hydrodynamic escape, because the higher molecular mass increases the temperature required to achieve a given value of $\lambda$. For example, to achieve $\lambda = 10$ on an Earth-like planet with an \ce{O}-dominated upper atmosphere given $R=1.3 R_\oplus$, $T \sim 9300$~K would be required. Whether or not such high temperatures can be reached depends on radiative and conductive effects in the thermosphere. Some gases (particularly \ce{CO2}) are predicted to cool the thermosphere effectively \citep{lichtenegger2010aeronomical}, although even \ce{CO2} may be lost hydrodynamically from small, hot planets \citep{tian2009thermal}. Uncertainties in the NLTE conditions \citep{lopezpuertas2001nonLTE} in the upper atmosphere currently limit our understanding of this regime. 

On smaller planets and moons, the weaker gravity means loss of heavy species like \ce{H2O} to space can occur much more readily \citep{arnscheidt2019atmospheric}. Finally, at extremely high temperatures, the definition of volatile shifts from species such as \ce{N2} and \ce{CO2} to Na, SiO and Mg. Escape of these species to space is expected on extremely hot rocky planets, and could explain the asymmetric transit profiles of some Kepler planet candidates \citep{rappaport2012possible,kang2021escaping}.

\begin{marginnote}
\entry{NLTE}{non-local thermal equilibrium}
\end{marginnote}

An additional class of heavy species escape processes exist that are non-thermal and hence not dependent on extreme heating of the thermosphere. Some non-thermal escape is still powered by stellar radiation. For example, photochemical escape occurs when photons break apart molecules or ions with sufficient excess energy that some of the products are moving fast enough to escape to space, in reactions such as 
\begin{equation}
\ce{O2+ + e- \to O + O}.
\end{equation}
Other non-thermal loss processes involve interaction with the stellar wind. Examples of important processes in the solar system include pick-up ion escape and sputtering \citep{lammer2008atmospheric}. The importance of both these effects depend strongly on the properties of the planet's magnetic field. It is sometimes stated that a magnetic field is required to protect a planet from atmospheric erosion, but this is a misconception: the terrestrial planet with the thickest atmosphere in the solar system (Venus) has no intrinsic magnetic field, and some simulations in fact predict an increase in escape rates as the dipole magnetic moment of terrestrial-type planets is increased \citep{gunell2018intrinsic}.

\begin{marginnote}
\entry{pick-up ion}{A neutral species from a planet's upper atmosphere that becomes ionized and picked up by the local magnetic field.}
\end{marginnote}

Complex three-dimensional models of non-thermal heavy species escape have now been applied to various exoplanets, including the planets of the TRAPPIST-1 system \citep[e.g., ][]{dong2018atmospheric,garcia2017magnetic}. The results of these models have suggested that high loss rates are possible from these planets, up to tens of bars of heavy species per Gy. However, many uncertainties remain in these estimates, because of unknowns in planetary magnetic fields, atmospheric composition, and the efficiency of loss processes.

Systematic differences in escape with host star type are somewhat easier to assess. Much work to date has focused on the differences between G-stars like the Sun and M-stars, because of the observational opportunities for planets around M-stars discussed in Section~\ref{sec:intro}. Three differences are particularly important. First, as noted in Section~\ref{subsec:atm_delivery}, M-dwarfs undergo a long pre-main sequence phase of elevated luminosity \citep[up to several 100~My for stars like TRAPPIST-1; ][]{baraffe2015new}. Second, many M-dwarfs emit considerably more XUV radiation over their lifetimes than G-dwarfs do, as a result of enhanced starspot activity and associated coronal heating \citep{france2016muscles, mcdonald2019xray}. Third, coronal mass ejection from M-dwarfs has also been predicted to be significantly greater, although observational constraints on this prediction remain limited \citep{crosley2018constraining,odert2020stellar}.

\begin{figure}
    \centering
    \includegraphics[width=5in]{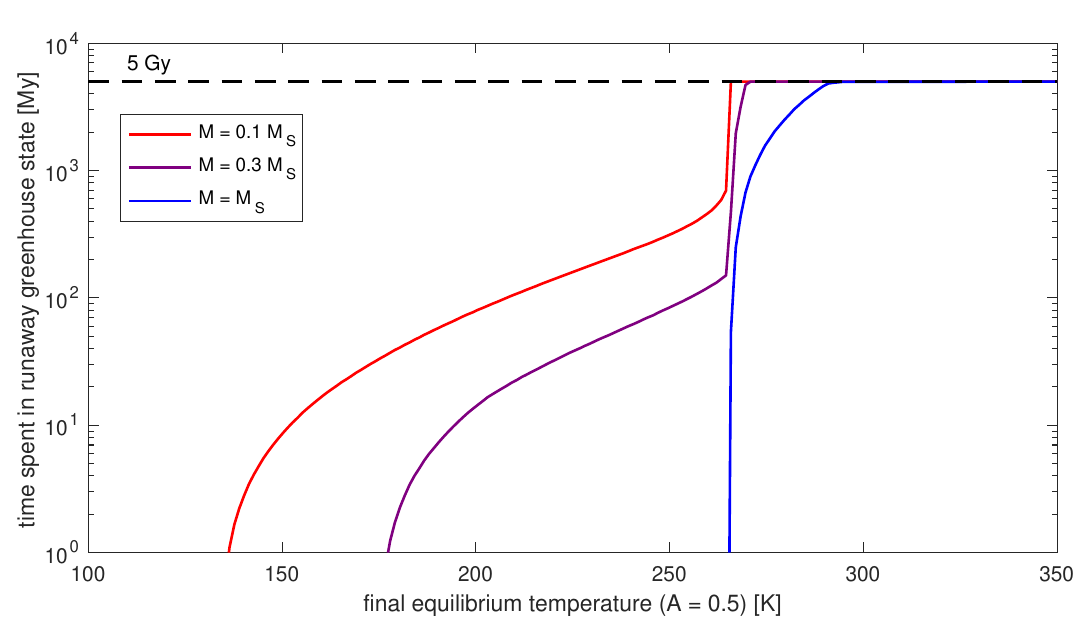}
    \caption{Total time spent in a runaway greenhouse state as a function of final equilibrium temperature, for rocky planets after 5~Gy of evolution. The extended pre-main sequence stage of low-mass stars causes planets that form around them to remain in a runaway greenhouse phase for up to several 100~My, leading to extensive water loss to space. Adapted from \cite{luger2015extreme}.  }
        \label{fig:lugerRG}
\end{figure}

These differences all imply enhanced atmospheric removal in M-dwarf systems. The high pre-main sequence luminosity will push many young rocky planets that obtain \ce{H2O} during formation into a runaway greenhouse state, where their upper atmospheres remain dominated by water vapor for long time periods (Figure~\ref{fig:lugerRG}). This \ce{H2O} is readily photolysed, and the high M-dwarf XUV fluxes can then power the escape of many Earth oceans worth of water to space, desiccating the planet \citep{ramirez2014habitable,tian2015water,luger2015extreme, bolmont2017water}. Because hydrodynamic drag preferentially removes hydrogen, this loss has important implications for atmospheric chemistry, as we discuss further in Section~\ref{sec:chemistry}.

A final major class of loss processes is atmospheric blowoff by meteorite impacts and large protoplanet collisions during formation. Meteorites can deliver volatiles, but the kinetic energy they supply on impact can also be a significant driver of atmospheric loss \citep{melosh1989impact}. This loss process is still incompletely understood, but it may have played a major role in Mars' atmospheric loss \citep{brain1998atmospheric}, and could also explain why Titan today has a thick \ce{N2} atmosphere, while similar outer solar system moons like Ganymede do not \citep{zahnle1992impact}. Smaller impactors may have a particularly important role in determining the differences in the noble gas inventories of Earth and Venus \citep{schlichting2015atmospheric}. However, few constraints on impactor fluxes to exoplanets over time currently exist.

Given all the uncertainties, empirical modeling of escape processes on rocky exoplanets is a compelling complementary approach. One example of such an approach is the cosmic shoreline proposed by \cite{zahnle2017cosmic}. \cite{zahnle2017cosmic} noted that the divide between airless bodies and those with atmospheres in the solar system follows a line $F\propto v_{esc}^4$, where $F$ is the present-day bolometric solar flux and $v_{esc}$ is the escape velocity for the body in question. Interestingly, the prediction from XUV-driven escape, which instead yields $F\propto v_{esc}^3$, conflicts with this empirical result. Data on the presence/absence of atmospheres for a wide range of exoplanets will be crucial to testing the generality of such scaling laws in the future (Section~\ref{sec:future}).

With future advances in instrumentation, it will become possible to test theories of escape directly. Lyman-$\alpha$ observations have already been used successfully to detect escaping hydrogen around gas-rich planets \citep{ehrenreich2015giant}, and have been attempted for rocky planets, though only upper limits have been obtained so far \citep{kislyakova2019transit, waalkes2019}. Detecting escape of heavier elements is another possibility, particularly the metastable helium absorption triplet, which is accessible with ground-based telescopes \citep{oklopvcic2018new, nortmann2018he}. Escaping ionized metals may also be detectable \citep{garcia2021pimen}, as well as auroral emissions from species such as oxygen \citep{luger2017pale}.

\section{ATMOSPHERIC CHEMISTRY}\label{sec:chemistry}

As noted earlier, the extreme diversity of rocky planet atmospheres relative to solar composition is an outcome of delivery and loss to space, internal chemistry, and exchange with the interior. In this section we discuss atmospheric chemistry and lay the groundwork for discussing interior exchange (which is itself strongly dependent on chemistry).

\subsection{Redox and equilibrium chemistry}

To classify atmospheric composition in a systematic way, it is useful to invoke broad chemical principles. For rocky planets, the most important of these by far is redox. The redox state of an system can be defined by comparing the total quantity of atoms like H or Fe, which readily donate electrons (reducing species) to those like O, which readily accept electrons (oxidizing species). In a planetary context, \ce{H2}- and \ce{CH4}-rich atmospheres such as Titan's are classified as highly reduced, \ce{CO2} or \ce{N2}-dominated atmospheres with trace \ce{H2} are weakly reduced, and atmospheres with an excess of \ce{O2} relative to reducing species, such as modern Earth's, are oxidized \citep{kasting1979oxygen,yung1999photochemistry}. Redox is closely tied to the topics of habitability and biogenesis, because reducing chemistry is necessary for prebiotic compounds like HCN to form in a planet's atmosphere \citep{ferris1978hcn,kasting1998setting}.

Seen through the lens of redox, hydrogen loss to space becomes a net oxidation of the planet.  The tendency of many rocky planets to oxidize can be viewed as a rather generic outcome of gravitational differentiation given the galactic abundance of elements, because H is lost to space and Fe sinks to the core, leaving an increased relative proportion of elements like O, C and N on the surface and in the atmosphere \citep{zahnle2013rise,wordsworth2018redox}. It is therefore reasonable to expect that the oxidation state of rocky exoplanet atmospheres will be correlated with their total amount of hydrogen loss. This oxidation can potentially continue to extreme levels: modeling suggests that in the absence of surface exchange, H loss from \ce{H2O} photolysis could lead to atmospheric \ce{O2} buildup on a planet without any biosphere \citep{wordsworth2014abiotic,luger2015extreme}. However, interaction with the surface and interior is an important additional process, as we discuss further in Section~\ref{sec:interiors}.

When an atmosphere's bulk elemental composition is known, equilibrium chemistry is the starting point for understanding which molecular species will be present. In pure equilibrium, atmospheric composition can be determined via minimization of Gibbs free energy and is solely a function of elemental abundance, temperature and pressure. At high temperatures, atmospheres are expected to be close to an equilibrium state, although disequilibrium processes may dominate on cooler planets, as we discuss in the next subsection. 

\begin{figure}[ht]
    \centering
    \includegraphics[width=5in]{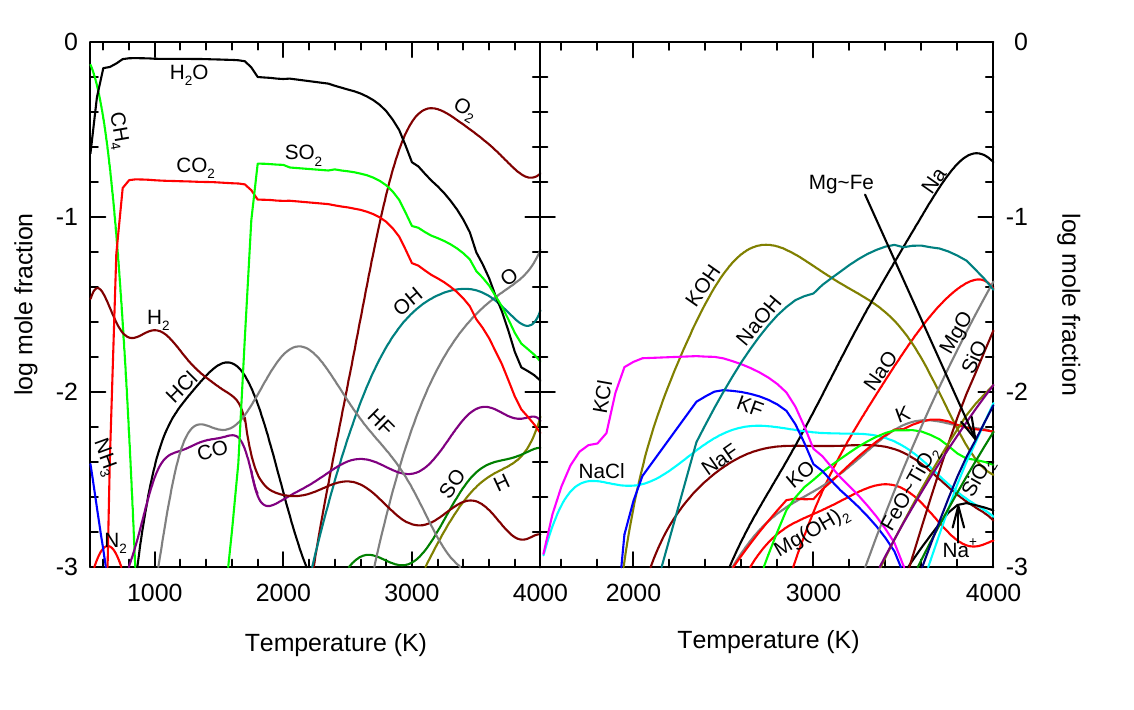}
    \caption{Hypothetical composition of a moderately oxidized exoplanet atmosphere assuming thermochemical equilibrium, BSE elemental abundances and 100~bars pressure. The left panel shows more volatile species, while the right panel shows more refractory species, and results are shown as a function of temperature. Adapted from \cite{schaefer2012vaporization}.}
    \label{fig:vaporization_earth}
\end{figure}

Equilibrium chemistry calculations and experiments for planets with temperatures in the 500 to 3500~K range predict a range of possible atmospheric compositions, depending on temperature, pressure and elemental abundance \citep{schaefer2009chemistry,miguel2011compositions,schaefer2012vaporization,thompson2021composition}. Under the reducing H-rich conditions expected for the most primitive atmospheres, \ce{H2}, \ce{CO} and \ce{CH4} are expected to dominate. When H, C, N, and S species are present but H is no longer a dominant species [e.g, for bulk silicate Earth (BSE) composition] \ce{H2O}, \ce{CO2}, \ce{SO2} and \ce{Na} are key atmospheric species  (Figure~\ref{fig:vaporization_earth}). Finally, for highly evolved atmospheres from which all H, C, N, and S has escaped, Na, \ce{O2}, O, SiO and Fe dominate atmospheric composition in the 2000 to 3500~K temperature range\footnote{These results assume that Na escapes less readily to space than C, N and S. In reality, it may also escape efficiently once atmospheric temperatures are high enough.}. These last results are relevant to several highly irradiated rocky exoplanets with $T_{eq}>500$~K (Table~\ref{tab:best_targets}). 

\begin{marginnote}
\entry{BSE}{Bulk Silicate Earth: The bulk chemical composition of Earth including the crust and mantle, but omitting the core.}
\end{marginnote}

\subsection{Photochemistry and other disequilibrium processes}\label{subsec:photochem}

Many rocky planets are cool enough for disequilibrium chemistry to play a major role in their atmospheric composition. On Earth, a key driver of chemical disequilibrium is life. Even on lifeless planets, however disequilibrium is caused by a host of processes, including volcanism, lightning, meteorite impacts, stellar coronal mass ejection events and, most importantly, photochemistry driven by stellar UV radiation \citep{airapetian2016prebiotic,ardaseva2017lightning,zahnle2020creation}.

Photochemical effects are most important in the upper atmosphere, where stellar UV radiation is absorbed. On gas giants, descending air parcels always reach a depth where thermochemistry drives them to local equilibrium, but for rocky planets, thermochemistry is only important if temperatures in the lowest part of the atmosphere are high enough. Venus is the only rocky planet in the solar system in this regime, but it may apply to many nearby rocky exoplanets. The shutdown of the drive to thermochemical equilibrium on planets with thin atmospheres could provide a way to probe exoplanet surface pressure indirectly in the future \citep{yu2021identify}, although, as we will see in the next section, interaction with the planet's interior can rarely be ignored.

Many photochemical modeling studies of rocky exoplanet atmospheres have focused on planets similar to Earth, although there is a steady trend towards more generalized models and conclusions. The problem of biosignature definition  around different star types (particularly M-dwarfs) fueled several early studies \citep[e.g., ][]{selsis2002signature,segura2003ozone,segura2007abiotic} and remains a key topic of interest.  Other modeling studies have addressed a wide range of topics, including differences between oxic and anoxic atmospheres \citep{hu2014photochemistry}, evolution of carbon and sulfur-bearing species \citep{hu2013photochemistry,hu2014photochemistry}, the formation of prebiotic compounds such as HCN \citep{rimmer2019hydrogen}, and three-dimensional coupling between chemistry and climate \citep[e.g., ][]{chen2019habitability, gomezleal2019climate}.

\begin{marginnote}
\entry{Far-ultraviolet (FUV)}{Electromagnetic radiation in the 122 to 200~nm wavelength range.}
\end{marginnote}

\begin{marginnote}
\entry{Near-ultraviolet (NUV)}{Electromagnetic radiation in the 300 to 400~nm wavelength range.}
\end{marginnote}

\begin{marginnote}
\entry{radical}{An atom, molecule or ion with an unpaired valence electron that is typically highly reactive.}
\end{marginnote}

One of the most important differences for photochemistry around stars like the Sun vs. lower-mass stars lies in the ultraviolet portion of their spectrum \citep{rugheimer2015uv}. Specifically, while K- and M-dwarfs emit a higher fraction of their luminosity in the XUV and FUV range, they emit a lower fraction of their luminosity as NUV and longer wavelength radiation because of their cooler photospheric temperatures. This leads to systematic differences in the ways that atmospheric chemical processes are expected to play out. One example of this is the photodissociation of \ce{CO2} by photons of wavelength $\lesssim 200$~nm 
\begin{equation}
\ce{CO2 + h\nu \to CO + O}\label{eq:CO2dissoc}
\end{equation}
In this reaction, the liberated O can readily form \ce{O2}, but reconversion of \ce{CO} and \ce{O} back to \ce{CO2} is inhibited by the slow rate of the spin-forbidden reaction \mbox{\ce{CO + O + M \to CO2 + M}}. As a result, catalytic cycles involving radical species such as OH are critical to determining the ratio of \ce{CO2} to \ce{CO} and \ce{O2}. However, the rate of OH production depends on photolysis of \ce{H2O} in many atmospheres. \ce{H2O} is more fragile than \ce{CO2} and hence can be dissociated by longer wavelength NUV radiation.

\cite{tian2014high} and \cite{gao2015stability} suggested that because M-stars have smaller NUV to FUV flux ratios than G-stars, photolysis of \ce{H2O} and hence production of \ce{OH} on Earth-like planets orbiting them would be limited, leading to enhanced atmospheric abundances of \ce{CO} and \ce{O2}. However, these results have recently been followed by new experimental measurements of the NUV \ce{H2O} photodissociation cross-section \citep{ranjan2020photochemistry} indicating lower \ce{O2} buildup due to reaction \eqref{eq:CO2dissoc} in some cases. The results of \cite{ranjan2020photochemistry} highlight the continued importance of performing laboratory measurements in tandem with theoretical modeling in planetary chemistry. 

To aid photochemical modeling in the future, accurate estimates of X-ray and UV fluxes from host stars are essential, and observational campaigns to develop comprehensive stellar UV databases \citep[e.g., ][]{shkolnik2014hazmat, france2016muscles} must continue. Future theoretical work must also pay greater attention to uncertainties in both absorption cross-sections and reaction rates. Monte Carlo photochemical simulations \citep[e.g., ][]{hebrard2005sensitivity} allow key sources of uncertainty to be isolated and could help identify the new laboratory experiments and ab-initio rate calculations that are most needed in the future.

\subsection{Aerosol formation: Cloud condensation and photochemical hazes}\label{sec:aerosol}

Aerosol formation due to condensation and heterogeneous chemistry is an extremely important and complex process in rocky planet atmospheres. Atmospheric aerosols may form purely due to condensation (e.g., \ce{H2O} or  \ce{CO2} clouds), or due to a combination of radical chemistry and condensation (e.g., \ce{H2SO4} aerosols and organic hazes). Clouds and aerosols have a major impact on observables and on planetary climate, because their radiative impact per unit mass is typically far greater than for species in the gas phase. Cloud and haze layers have been identified in the atmospheres of several hot Jupiter and sub-Neptune exoplanets \citep{kreidberg2014clouds, knutson2014featureless, sing2016continuum}, and there is every reason to expect they will be equally important on rocky exoplanets.

\begin{marginnote}
\entry{aerosol}{A general term for any solid or liquid particle (cloud, haze or dust) suspended in a gas.}
\end{marginnote}

Photochemical hazes are a particularly important class of aerosols that occur primarily in reducing atmospheres rich in hydrogenated C species such as \ce{CH4}. Photolysis of \ce{CH4} produces radicals such as \ce{CH3}, which can combine to form longer chain species and eventually condense to create haze particles. Photochemical hydrocarbon hazes are observed on many outer solar system bodies, including most famously Titan. They are also present in the atmospheres of Jupiter and Saturn, although at lower abundance, because reaction of \ce{CH3} and other radicals with \ce{H2} is effective at inhibiting the buildup of \ce{C_nH_n} chain species \citep{yung1999photochemistry}. Once they form, it is common for haze particles to aggregate into fluffy fractal particles that have unique radiative and microphysical properties \citep{rannou2003model}. Photochemical hazes are also thought to have been present on the early Earth, before the rise of oxygen \citep{domagal2008organic,zerkle2012bistable}, and may play an important role in the climates and observable properties of many rocky exoplanets \citep{arney2017pale}.

The incorporation of other elements such as N and O into haze particles depends on their abundance and reactivity. The \ce{N2} bond is notoriously hard to break, but it can be dissociated by XUV radiation and energetic particles in the upper atmosphere or lightning in the lower atmosphere  \citep{yung1999photochemistry,rimmer2019hydrogen}. N atoms can rapidly react with a number of species, including hydrocarbon radicals to form the quintessential prebiotic feedstock molecule, HCN. Oxygen is a minor contributor to haze particles on the outer solar system planets and satellites because of \ce{H2O} cold-trapping, but it is likely to be important on a number of rocky exoplanets. Broadly speaking, haze formation is disfavored in atmospheres with low C/O ratios \citep[below about 0.6; ][]{trainer2006organic}, but some production can still occur even when \ce{O2} is present \citep{ugelow2018effect}, as any resident of a polluted city on Earth knows.

Sulfur-rich hazes can form under both reducing and oxidizing chemical conditions. Under reducing conditions, sulfur haze particles may consist of elemental sulfur (\ce{S8}) or organosulfates \citep{dewitt2010formation}. Under oxidizing conditions, photolysis of \ce{SO2} and/or combination with OH radicals followed by condensation in the presence of \ce{H2O} leads to \ce{H2SO4} aerosols. These aerosols can dramatically increase a planet's albedo even in an atmosphere with low \ce{H2O} and \ce{SO2} abundance, as is the case on present-day Venus. Because \ce{H2SO4} particles slowly sediment and are highly soluble in liquid water, the presence of a long-lived high-albedo \ce{H2SO4} cloud deck is incompatible with surface liquid water \citep{loftus2019sulfate}, and probably also with a thin, non-thermolyzing atmosphere (see also Section~\ref{sec:interiors}).

Current understanding of other possible aerosol types on rocky exoplanets is limited, but growing. For example, \cite{mbarek2016clouds} performed a general study of chemical equilibrium condensate species in rocky and sub-Neptune exoplanet atmospheres for various compositions and temperatures ranging from 350 to 3000~K. They predicted a wide variety of cloud types, including exotic possibilities such as \ce{K2SO4}, \ce{ZnO} and graphite. Laboratory aerosol experiments also play an essential role in testing theoretical calculations and exploring the range of possibilities for exoplanet atmospheres.  For example, \cite{horst2018haze} performed cold plasma discharge experiments and found haze production that was more diverse and widespread than predicted by theory. \cite{he2020sulfur} expanded this investigation to sulfur species in \ce{CO2}-rich atmospheres, and found an important role for \ce{H2S} in production of complex sulfur products.  Future progress in this area will require detailed intercomparison between laboratory results and numerical chemical kinetic models.

\section{ATMOSPHERE-INTERIOR INTERACTION}\label{sec:interiors}

By definition, rocky planet atmospheres constitute a small mass fraction of the planets they envelop. As a result, their compositions are often strongly dependent on exchange with the interior. Here, we discuss atmosphere exchange with several types of interior expected to be common on rocky exoplanets. 

\subsection{Exchange with magma oceans}

Put simply, a magma ocean is a region of a rocky mantle that has reached temperatures sufficient to melt it. Rock compositions vary in the solar system (and will likely vary even more for exoplanets), but for a typical peridotitic composition, the low-pressure solidus temperature is around 1400~K, while the liquidus is around 2000~K \citep{hirschmann2000mantle}. Between these temperatures, a well-mixed magma consists of a slurry-like mixture of solid and liquid phases. With the possible exception of Io, whose interior is heated by tidal interactions with Jupiter \citep{khurana2011evidence}, silicate magma oceans are not observed in the solar system today. However, a number of lines of evidence indicate they played a crucial role in atmospheric evolution in the early solar system, during and just after formation of the terrestrial planets \citep{elkins2012magma}. Transient magma oceans are probably common on young rocky planets, and many hot rocky exoplanets may have permanently molten surfaces, particularly if they also possess thick atmospheres.

\begin{marginnote}
\entry{peridotite}{The dominant rock type in Earth's mantle, composed of a mixture of \ce{SiO2}, MgO and FeO.}
\end{marginnote}

\begin{figure}[ht]
    \centering
    \includegraphics[width=5in]{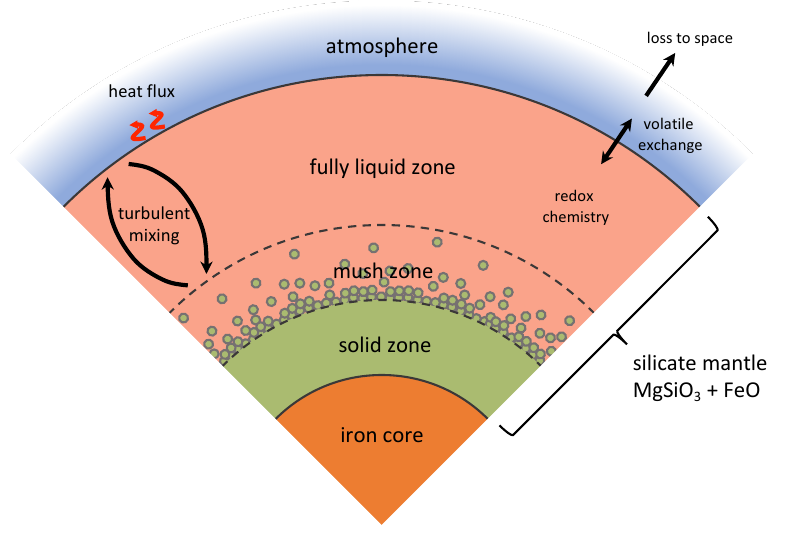}
    \caption{Schematic of a rocky planet interior with a magma ocean, indicating the solid-liquid transition with radius alongside key physical and chemical processes. Partially adapted from \cite{lebrun2013thermal}.}
    \label{fig:magma_ocean_schematic}
\end{figure}

The mean depth of a magma ocean depends primarily on surface temperature. Convection in most magma oceans is expected to be intense enough for rapid exchange of volatiles with the atmosphere and an adiabatic vertical temperature structure to occur. The intersection of the magma adiabat with the solidus in pressure-temperature space is such that magma oceans should freeze from the bottom upwards  \citep{elkins2012magma,lebrun2013thermal}. Importantly, this means that magma oceans should continue to efficiently exchange volatiles with the overlying atmosphere until they solidify at the surface (Fig.~\ref{fig:magma_ocean_schematic}).

In the absence of external heating, airless planets with magma oceans cool rapidly: given a magma heat capacity of 1200~J~kg$^{-1}$~K$^{-1}$, density of 3000~kg~m$^{-3}$ and flux to space $F=\sigma T^4$ (with $\sigma$ Stefan's constant and $T$ temperature), the timescale to cool a 100-km deep magma ocean from the liquidus temperature to the solidus is of order $\tau_{cool} \sim 15$~years. 
However, magma ocean cooling is dramatically slowed or even halted by the presence of an atmosphere, because of absorption of upwelling IR radiation (i.e., the greenhouse effect). 

Atmospheric IR opacity depends on composition. Because \ce{H2} is a particularly potent greenhouse gas at pressures above $\sim 0.1$~bar (Section~\ref{sec:insights}), rocky planets that form with even thin hydrogen envelopes will initially have completely molten surfaces \citep{hayashi1979earths,chachan2018role}. \ce{H2O} is also a strong greenhouse gas, but it condenses when a planet's equilibrium temperature is below the runaway greenhouse limit. This may have been important to the divergent evolution of Venus and Earth, because Earth's received solar flux was low enough to allow condensation of water after a few My, while Venus would have retained a steam atmosphere until most of its \ce{H2O} was destroyed by photolysis followed by H loss to space \citep{hamano2013emergence}. This divergence based on received stellar flux may also be important to the atmospheric evolution of many exoplanets \citep{kane2019venus}.

Magma oceans affect atmospheric composition because atmospheric volatiles dissolve in them, and sometimes also interact with them chemically. The partitioning of gases into a magma depends strongly on its redox state  \citep[e.g., ][]{gaillard2022redox}. The most important determinant of redox in a silicate mantle is iron, because of its abundance and ability to transition between the \ce{Fe^{0}}, \ce{Fe^{2+}}  and \ce{Fe^{3+}} states (Mg is also abundant in the mantle, but it is nearly always found in the +2 oxidation state). A high abundance of \ce{Fe^{0}} and \ce{Fe^{2+}} corresponds to highly reducing conditions, while the \ce{Fe^{3+}} oxidation state is abundant only in highly oxidizing conditions. 

Iron segregation to a rocky planet's core during formation is never 100\% efficient, as it depends on factors such as FeO equilibrium chemistry during core formation \citep{fischer2011equation}. Based on modeling and analogy with the solar system, mantle iron contents of a few to 10~wt\% or more for rocky exoplanets are expected. This translates to a vast chemical reservoir. As an example, 
if all the iron in Earth's mantle \citep[around 6~wt\%; ][]{javoy1999chemical} were in the form of \ce{Fe^{2+}}, conversion of that iron to \ce{Fe^{3+}} through the reaction \ce{2FeO + O \to Fe2O3} could consume the O equivalent to around 28~Earth oceans  ($2.2\times 10^{24}$~mol). 
This means that if a magma ocean is present on planets that are oxidizing through \ce{H2O} photolysis and H loss, it can absorb huge amounts of O before \ce{O2} began to build up in the atmosphere.

\begin{marginnote}
\entry{wt\%}{Ratio of element mass to total mass, expressed as a percentage.}
\end{marginnote}

Geochemists usually express mantle redox state in terms of various mineral redox buffers. In terms of iron, this translates to differing proportions of \ce{Fe^{0}}, \ce{Fe^{2+}} and \ce{Fe^{3+}}. 
Earth's upper mantle today is close to the quartz-fayalite-magnetite (QFM) buffer, with \ce{Fe^{3+}}/\ce{Fe_{tot}}$\sim$~0.03  \citep{armstrong2019deep} (here \ce{Fe_{tot}} is the total iron abundance of the mantle). During Earth's formation, however, differentiation was ongoing and much more metallic iron (\ce{Fe^{0}}) was present. The mantle was hence much more reducing, probably below the iron-w\"ustite (IW) buffer, with an \ce{Fe^{3+}}/\ce{Fe_{tot}} ratio of around 0.004 \citep{armstrong2019deep}.  Earth's increase in mantle oxidation over time may in part be due to early oxidation of the atmosphere via hydrogen loss. However, modern deuterium to hydrogen (D/H) ratios in terrestrial seawater are quite close to the value for chondritic meteorites. This observation suggests that Earth's total loss of hydrogen from \ce{H2O} photolysis was relatively modest  \citep[e.g., ][]{zahnle2020creation}, probably because of effective cold-trapping of water in the lower atmosphere. Venus, in contrast, may have a highly oxidizing mantle closer to the magnetite-hematite (MH) buffer due to extensive early H loss, although more data on its surface geochemistry is needed \citep{fegley1997oxidation,wordsworth2016atmospheric}.

\begin{marginnote}
\entry{redox buffer}{A collection of minerals that together define the chemical state of a silicate mantle region.}
\end{marginnote}

Another process, Fe redox disproportionation, may also have played a role in mantle redox evolution on Earth \citep{armstrong2019deep}, and is likely important on exoplanets. In this process, iron in the 2+ oxidation state disproportionates to \ce{Fe^{3+}} and metallic iron (\ce{Fe^{0}}):
\begin{equation}
\ce{3Fe^{2+} \to Fe^0 + 2Fe^{3+}}.
\end{equation}
Because the density of metallic iron is greater than that of mantle minerals, it sinks to the core. After core formation is complete, the end result is a more oxidized mantle. Redox disproportionation is yet another example of the deep connection between gravitational differentiation and chemical evolution on rocky planets. It is most efficient at high pressure, so planets of Earth mass and greater are expected to develop oxidized mantles more efficiently than smaller planets such as Mars \citep{deng2020magma}.

Because the atmosphere and interior exchange volatiles rapidly when a magma ocean is present, the mantle redox state is a key determinant of atmospheric composition in this case. A reducing magma ocean near the IW buffer would outgas species like \ce{CO} and \ce{H2}. Conversely, a planet with enough $\ce{Fe^{3+}}$ to place it above the magnetite-hematite (MH) buffer would be so oxidized that \ce{O2} itself would become an important volatile outgassed to the atmosphere.

Mantle redox also determines the solubility of key volatile species such as C and N in magma. Under oxidizing conditions, \ce{CO2} is the chemically favored form of dissolved carbon in magma oceans. The solubility of \ce{CO2} in magma is low, which means C-rich oxidized magma ocean planets are generally expected to have dense \ce{CO2} atmospheres \citep[e.g., ][]{lebrun2013thermal}. However, in a reducing magma, C forms C-O and C-H species, Fe-carbonyl complexes, and even graphite, greatly increasing solubility and hence limiting the atmospheric \ce{CO2} abundance \citep{grewal2020speciation,fischer2020carbon}.
The behaviour of N is similar: \ce{N2} is quite insoluble in oxidized magmas, but it has greatly enhanced solubility under more reducing conditions \citep{libourel2003nitrogen,grewal2020speciation}. This redox dependence of interior sequestration implies that rocky planets with reducing mantles should have relatively C- and N- poor atmospheres. This may have important implications for future observations.

The theoretical understanding of magma ocean interaction with atmospheres on rocky exoplanets has evolved rapidly over the last decade. Several studies have now analyzed the changes in atmospheric composition resulting from exchange with magma oceans of different compositions, volatile abundances and redox states \citep{lupu2014atmospheres,kite2016atmosphere,katyal2020effect,lichtenberg2021vertically}. 
The development of models that can track the evolving redox chemistry of the atmosphere -- magma ocean system as hydrogen is progressively lost to space \citep{schaefer2016predictions,wordsworth2018redox,krissansen2021oxygen} has also been an important step forward. Notably, these studies have demonstrated that while magma oceans are huge sinks for oxygen liberated from water loss, there are still many situations where \ce{O2}-dominated atmospheres can appear abiotically, particularly on hot planets. 

Testing the predictions of coupled atmosphere-magma ocean models observationally is an important future goal. While many processes contribute to atmospheric evolution, modeling suggests that the oxidation state of a planet's atmosphere should be correlated with the planet's radius, orbital distance and host star type, with close-in planets around M-type stars particularly likely to have oxidized atmospheres. Theoretical predictions for the most accessible targets listed in Table~\ref{tab:best_targets} are discussed further in Section~\ref{sec:future}. 

\subsection{Exchange with water and other liquids}

\ce{H2O} is cosmically abundant and central to the concept of habitability (Section~\ref{sec:future}), so water oceans are another important type of liquid surface to consider on low-mass exoplanets. Many theoretical studies on this topic have focused on rocky planets that have such high surface \ce{H2O} inventories that they no longer have any exposed land (waterworlds). Variability in \ce{H2O} delivery and loss (Section~\ref{sec:atm_escape}) means that such planets may be common, and initial modeling suggests that their atmospheres and climates should be dramatically differently from Earth's. 

The solubility of chemical species in \ce{H2O} varies widely: while non-polar species such as \ce{N2} and \ce{H2} are relatively insoluble, others (\ce{CO2}, \ce{NH3}, \ce{SO2}) are readily dissolved  \citep[e.g., ][]{pierrehumbert2010principles,abbot2012indication}. 
The solubility of \ce{CO2} in particular decreases with temperature in the 300 to 400~K temperature range \citep{carroll1991solubility}. This can lead to destabilizing feedbacks, because more outgassing implies warmer surface temperatures, which in turn causes still more outgassing \citep{wordsworth2013water,kitzmann2015unstable}. Various proposals have been put forward to allow \ce{CO2} levels to remain at the right level in the atmospheres of waterworlds to maintain habitable temperatures \citep{levi2017abundance,kite2018habitability,ramirez2018ice}. However, for many volatile-rich planets, it is plausible that enough \ce{CO2} could be acquired during formation to prevent the surface condensation of water to form oceans in the first place \citep{marounina2020internal}. The question of waterworld habitability is discussed further in Section~\ref{sec:future}.

The solubility of other volatile species in water may provide opportunities to identify liquid water oceans on exoplanets. For example, \cite{loftus2019sulfate} showed that the high solubility of oxidized sulfur in water should limit atmospheric abundances of \ce{SO2} and \ce{H2SO4} to undetectable levels on planets with even small surface oceans \citep[although transient \ce{SO2} can still be present as a result of volcanism; ][]{kaltenegger2009detecting}. \cite{hu2021unveiling} recently proposed that a similar approach could be applied using \ce{NH3} to identify water oceans on rocky planets with \ce{H2}-dominated atmospheres. However, N is soluble in the reducing magmas that are expected when \ce{H2} is the dominant atmospheric gas, so it is unclear if such observations could distinguish water oceans from magma ones.

Comparatively little research has been done on interior exchange involving other liquids. Photochemical haze formation (Section~\ref{sec:aerosol}) can lead to formation of hydrocarbon lakes, as on Titan \citep{mitri2007hydrocarbon}, or mixtures of even more complex organics, as may have happened on the early Earth \citep{arney2017pale}. On a planet with a moderately or strongly reducing atmosphere, thick layers of liquid hydrocarbons or more complex organics could build up on the surface over time in the absence of strong sinks. The implications of hydrocarbon oceans or other exotic possibilities for atmospheric observables have not yet been studied in any detail.

\subsection{Solid surface exchange processes}

When planetary surfaces solidify, the rate of volatile exchange with the atmosphere changes dramatically.
The internal mixing timescales of oceans (magma, water or otherwise) are typically very fast compared to atmospheric loss and supply timescales, so equilibrium with the atmosphere can usually be assumed. In contrast, the mixing timescales of solid interiors are orders of magnitude slower \citep[e.g., $\sim$1~Gy for He in Earth's mantle; ][]{coltice2006mixing}. Furthermore, there is still great uncertainty in how mantle convection and tectonics evolve on rocky planets through their lifetimes \citep{hawkesworth2018earth, noack2014tectonics}. For all these reasons, predicting solid surface exchange processes for exoplanets is a huge challenge.

Volcanic outgassing is the primary process by which volatiles are released from Earth's interior. The rate of volcanic outgassing depends on the rate of crustal production, which itself is a function of the tectonic regime of the planet. On a planet in a plate-tectonic regime, such as present-day Earth, volatile evolution models can be constructed that track the outgassing of species like \ce{H2O} over time using scaling parameters from convection theory \citep[e.g., ][]{schaefer2015persistence}. Certain broad trends are predicted from such models: for example, many young planets with high mantle potential temperature should experience fast crustal growth and high rates of volcanic outgassing \citep{honing2019carbon}. 

The scaling of tectonic regime with planet mass is less well understood. Early work suggested that plate tectonics becomes more likely as a planet's mass increases \citep{valencia2007inevitability}, but this has since been contested. Subsequent studies have suggested that factors such as the presence or absence of \ce{H2O} \citep{korenaga2010likelihood} or the planet's thermal history \citep{lenardic2012notion} may be more important than its mass. For planets in the stagnant-lid regime, outgassing rates are expected to decrease with increasing planet mass \citep{noack2017lid}. Given the uncertainties about whether plate tectonics will occur, a wide range of possible outgassing rates should be expected. 

What solid interiors provide to atmospheres via outgassing they can also take away, because atmospheric volatiles react chemically with planetary crusts. As with other chemical processes, the reaction rates are strong functions of temperature, but they also depend on the presence of solvents, particularly liquid \ce{H2O}. Carbonate formation is a classic example of a process that is thermodynamically favored at Earth-like temperatures, but dependent on the presence of liquid water to proceed at geologically important rates. In this process, Ca and Mg cations produced from the weathering of fresh igneous rocks react in seawater with bicarbonate ions 
\begin{equation}
\ce{Ca^{2+}} + 2\ce{HCO3^-} \to \ce{CaCO3} + \ce{CO2} + \ce{H2O}
\end{equation}
removing net \ce{CO2} from the atmosphere-ocean system in the process. On Earth today, the rate of carbonate formation is sufficient to completely remove \ce{CO2} from the atmosphere-ocean system on a timescale of under a million years \citep{berner2004phanerozoic}. 
This does not happen because removal of \ce{CO2} into carbonates is balanced by volcanic outgassing. A negative feedback, the carbonate-silicate cycle, is believed to keep volcanic outgassing and carbonate formation closely balanced on long timescales on Earth \citep[][ see also Section~\ref{sec:future}]{walker1981negative}.

A large number of additional reactions between volatiles and crustal material are important to rocky planet atmospheric evolution. Phyllosilicate formation due to interaction of igneous rocks with water can constitute a significant \ce{H2O} sink to the crust and/or mantle over time. Serpentines are particularly interesting from an atmospheric standpoint, because reduced gases such as \ce{H2} and \ce{CH4} may be produced as a byproduct of their formation, which can have many implications for chemistry and climate \citep{oze2005have,etiope2013low,wordsworth2017transient}. Aqueous chemistry can also facilitate redox reactions, particularly the oxidation of iron by UV radiation or atmospheric \ce{O2} \citep{klein2005some,hurowitz2010origin}. This is particularly important to the question of when \ce{O2} can be regarded as a biosignature on Earth-like exoplanets (Section~\ref{sec:future}).

\begin{marginnote}
\entry{phyllosilicate}{Hydrated silicate minerals, including clays, micas and serpentines, that form in parallel sheets of silicate bonded with water or hydroxyl (OH) groups.}
\end{marginnote}

For planets that do not possess surface liquid water, dry atmosphere-crust reactions can still be important if surface temperatures are high enough, as is the case on Venus in the solar system today \citep{zolotov2018gas}. On rocky exoplanets with surface temperature in the 500 to 1400~K range, this class of reactions is also likely to be important, although our understanding of the range of possibilities is currently poor. For planets with solid compositions that have no analogue in the solar system (e.g., carbon-rich worlds), we currently have only limited understanding of the range of possible tectonic regimes or atmosphere-interior interaction chemistries, although major differences from the solar system planets appear plausible \citep{unterborn2014role,hakim2019thermal}. Again, more theoretical and laboratory analysis will be required here to help constrain atmospheric evolution models in the future.

\section{ATMOSPHERIC DYNAMICS AND CLIMATE}\label{sec:atm_dyn}

All planetary atmospheres are in a state of continuous motion, as a result of differential heating and rotation. In contrast to gas giant planets, where heat fluxes from the interior can be important, atmospheric motion on the vast majority of rocky planets is driven by stellar radiation. Atmospheric dynamics determines wind speeds, which are potentially observable via high-resolution Doppler spectrometry, and heat transport from hot to cold regions of the surface, which leaves observable signatures in thermal phase curves (Section~\ref{sec:observations}). It also influences atmospheric composition and cloud formation by determining the distribution of condensable species such as \ce{H2O}. This in turn affects a planet's potential habitability.

Atmospheric dynamics is studied theoretically via a range of tools, from pen and paper analytics to global circulation models (GCMs) that solve the fluid dynamical equations of a planet's atmosphere in 3D. A number of studies have now systematically investigated how atmospheric dynamics is likely to change as a function of planet mass, rotation rate, stellar distance, atmospheric shortwave opacity, and other factors \citep[e.g.,][]{showman2013atmospheric,carone2014,carone2015,carone2016, kaspi2015atmospheric,wang2018comparative,read2018comparative,komacek2019scaling,kang2019collapse,auclair2020atmospheric}. Broadly speaking, these studies have shown that the temperature variations across a rocky planet's surface increase with planetary radius and rotation rate, and decrease with atmospheric pressure. Wind speeds also vary greatly, with equatorial super-rotation often occurring on planets with slow rotation rates or elevated visible wavelength atmospheric opacity. 

\begin{marginnote}
\entry{equatorial super-rotation}{Dynamical regime where the angular momentum of the atmosphere is greater than that of the solid planet near the equator.}
\end{marginnote}

\begin{figure}[ht]
    \centering
    \includegraphics[width=5in]{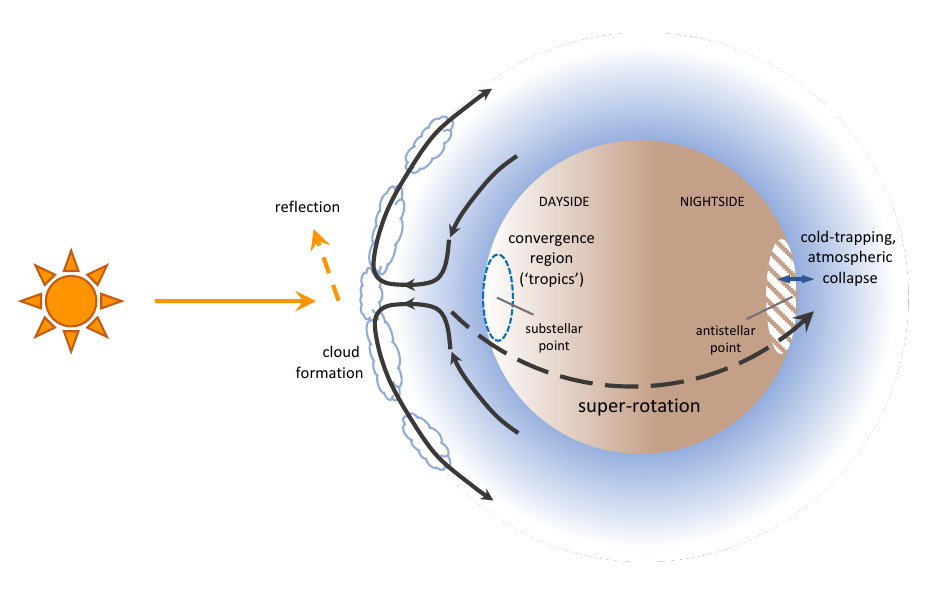}
    \caption{Key dynamical and climate processes in the atmosphere of a tidally locked rocky exoplanet. The hydrological cycle depends on the total \ce{H2O} inventory and other factors: cloud coverage and surface inventories are shown here for a planet with low total \ce{H2O} compared to Earth.}
    \label{fig:circulation_schematic}
\end{figure}

\subsection{Atmospheric circulation of tidally locked planets}

The atmospheric circulation of rocky planets around M-dwarf stars has received special theoretical attention over the last decade. Because M-dwarfs are dim compared to the Sun, planets around them that receive similar stellar fluxes to Earth are in  close orbits. For example, Proxima Centuri b, which receives 65\% of the flux of Earth, has a semi-major axis of 0.0485~AU and a year length of just 11.2 days \citep{anglada2016proxima}.
For such small star-planet separations, tidal effects are extremely important. Hence many M-dwarf rocky planets are predicted to have low orbital obliquities and tidally resonant orbits \citep{heller2011tidal}, although on some planets, asynchrony may be maintained by atmospheric thermal tides \citep{leconte2015asynchronous}.  The most extreme case of tidal interaction is a 1:1 resonance, or tidal locking. A planet in this configuration has permanent day and night sides, which profoundly affects its atmospheric circulation and climate. 

Theory and modeling predict that a planet with permanent day and night sides will develop strong convergence and upwelling around the substellar point, horizontal transport from dayside to nightside, and a return flow to the dayside near the surface, in a kind of global convection cell (Figure~\ref{fig:circulation_schematic}). Planetary rotation breaks the radial symmetry of this flow along the sub-stellar/anti-stellar axis, which forces the development of Rossby and equatorial Kelvin waves  
\citep{merlis2010atmospheric,showman2010matsuno}. The resulting flow pattern then drives angular momentum toward the equator, leading to equatorial super-rotation in many cases \citep{merlis2010atmospheric,showman2011equatorial}. On planets with thick atmospheres, the interaction of the deep and upper atmospheres makes predicting equilibrated super-rotating flows a significant challenge \citep{wang2020extremely,carone2020equatorial}, but for rocky exoplanets with thinner atmospheres, flow equilibration times are much shorter and model predictions are in principle more reliable. Direct measurement of the wind speed in a rocky exoplanet atmosphere remains a key future observational goal (Section~\ref{sec:newobs}).

\begin{marginnote}
\entry{equatorial Kelvin wave}{Planetary-scale atmospheric wave that propagates eastward along the equator on a rotating planet.}
\end{marginnote}

\begin{marginnote}
\entry{Rossby wave}{Planetary-scale atmospheric wave that propagates westward at mid-latitudes on a rotating planet.}
\end{marginnote}

The horizontal heat transport by atmospheres on tidally locked planets is particularly important, because of its effect on thermal phase curves and volatile transport. On fast rotating planets, the presence of eddies and turbulence make this heat transport challenging to predict, although scaling theories have been developed \citep{komacek2019scaling}. On slower rotators, the weak temperature gradient (WTG) approximation can be made, which greatly aids analysis \citep{merlis2010atmospheric,pierrehumbert2010palette,wordsworth2015atmospheric,koll2016temperature}. An atmosphere in a WTG regime can be treated as isothermal above the planetary boundary layer, so the interhemispheric heat transport problem boils down to determining the enthalpy fluxes (sensible, radiative and latent) between the surface and atmosphere in each hemisphere \citep{wordsworth2015atmospheric,koll2016temperature,auclair2020atmospheric}. Both 3D modeling and theory predict that nightside warming strongly increases with both the pressure and the IR opacity of the atmosphere. For condensable species like \ce{CO2}, the pressure dependence can lead to a runaway process called atmospheric collapse (Section~\ref{sec:coldtrap}).

Atmospheric heat transport is one of the few processes that has already been tested observationally on a rocky exoplanet. Using the Infrared Array Camera (IRAC) on Spitzer, \cite{kreidberg2019absence} performed thermal phase curve observations of the hot rocky exoplanet LHS~3844b (see also Section~\ref{sec:observations}). They found a symmetric phase curve and a high dayside flux, implying inefficient day to nightside  heat transport and hence a thin atmosphere. Via detailed comparison with theory, they were able to rule out the presence of an atmosphere with surface pressure above 10 bar. Thermal phase curve analysis is expected to be a key tool for atmospheric characterization of rocky exoplanets in the future, particularly when combined with spectral analysis \citep{selsis2011thermal}.

\subsection{Cold-trapping, atmospheric collapse, and ice-albedo feedbacks}\label{sec:coldtrap}

Condensation is fundamental to determining the atmospheric composition of rocky planets. When horizontal heat transport is ineffective, some regions of the surface can become much, much colder than the planet's equilibrium temperature. These regions then act as cold traps for volatile species. If  the cold trap temperature $T_c$ is known, predicting the atmospheric abundance of a cold-trapped species is simple: in the absence of any additional sinks or sources in the atmosphere, it is uniquely determined by the saturation vapor pressure $p_{sat}(T_c)$  \citep[e.g.,][]{leconte20133d,ding2020stabilization}.

On rocky bodies in the solar system, permanently shadowed regions can be incredibly effective cold traps. Even Mercury, which receives almost seven times the solar flux of Earth, 
exhibits evidence of water ice and organic deposits in polar craters that are permanently shielded from sunlight \citep{neumann2013bright}. GCM simulations predict that for a tidally locked exoplanet with Earth's atmospheric composition, water would rapidly become trapped as ice on the nightside  \citep{leconte20133d}. However, when dayside tropopause temperatures are cold enough, an oasis regime can result where some water remains trapped as liquid on the planet's dayside \citep{ding2020stabilization}.

Atmospheric collapse is the extreme limit of volatile cold-trapping when the main component of the atmosphere is the species that condenses. Atmospheric collapse may sound dramatic, but it is actually common in our solar system: Mars, Triton and Pluto are all in collapsed atmosphere regimes. Once collapse starts, it is usually a runaway process, because thinner atmospheres transport heat and warm the surface less effectively, making cold-trap regions even colder still. For gases like \ce{H2O}, collapse is the direct inverse of the runaway greenhouse process mentioned in Section~\ref{sec:insights}.

In the 1990s, climate researchers initially thought atmospheric collapse might be fatal to the habitability of planets around M-stars, and so it was one of the first exoplanet questions to be studied using a 3D climate model \citep{joshi1997simulations,joshi2003climate}. Recent GCM simulations with real-gas radiative transfer predict that the collapse pressure for pure \ce{CO2} is a function of stellar flux and planet mass, with values from around 0.1~bar to 10~bar in the 1-10~$M_E$ and 500-4000~W/m$^2$ range \citep{wordsworth2015atmospheric,turbet2016habitability}. 
Very hot rocky exoplanets for which Na, SiO and Mg are the main volatile species are also expected to be in collapse regimes. Simplified circulation modeling suggests that flows on such planets will be supersonic from dayside to nightside, analogous to the \ce{SO2} circulation regime on Io \citep{castan2011atmospheres,kang2021escaping}. 

When volatiles condense across wide regions of a planet's surface, they can alter albedo and hence surface temperature, causing climate feedbacks. Ice-albedo feedbacks involving \ce{H2O} were responsible for the Snowball glaciations that occurred in Earth's past \citep{hoffman1998neoproterozoic}. On exoplanets, the frequency of Snowball transitions likely depends on factors such as the mean rate of volcanic outgassing and the internal variability of the carbon cycle \citep{haqq2016limit,wordsworth2021likely}. However, around M-dwarfs, the redder stellar spectrum weakens the feedback by decreasing ice albedo \citep{joshi2012suppression,shields2013effect} and tidal locking may inhibit Snowball bifurcations entirely \citep{checlair2017no,checlair2019no}. 

\subsection{Clouds and climate}

Clouds are important because they affect observables and climate. They are also a huge challenge to model. 
Once clouds condense, they can scatter and/or absorb both incoming visible and outgoing IR radiation, depending on their composition and size distribution. Species that absorb in the UV, visible or near-IR, such as photochemical hazes and ozone (\ce{O3}), can cause strong temperature inversions in the atmosphere and an anti-greenhouse effect that lowers surface temperatures \citep{mckay1991greenhouse}. However, many condensates, including \ce{H2O} and \ce{CO2} clouds, mainly scatter visible radiation. The net climate effect of these species then comes down to a competition between visible scattering and IR extinction effects, which leads to variations in the sign of radiative forcing with cloud altitude \citep{pierrehumbert2010principles}.

Multiple studies have now modeled radiatively active clouds on rocky exoplanets using GCMs \citep[e.g., ][]{wordsworth2011gliese,yang2013stabilizing,kumar2016inner,turbet2016habitability,wolf2017assessing,sergeev2020atmospheric,lefevre20213d}. Among other things, these studies have shown that \ce{CO2} clouds can cause warming on cool planets with thick \ce{CO2} atmospheres via IR scattering, supporting earlier predictions from 1D models \citep{forget1997warming}. Beginning with \cite{yang2013stabilizing}, they have also suggested that on slowly rotating planets, concentrated water clouds can cause enhanced reflection of starlight around the substellar point. This could help planets around M-stars maintain surface liquid water at higher incident stellar fluxes than planets around G-stars. 

In all cases, the overall magnitude of cloud effects is dependent on details of cloud microphysics and convection, which are usually poorly constrained. First-principles studies that focus on the behavior of convection \citep[e.g., ][]{ding2016convection,ding2018global} or cloud and precipitation microphysics \citep[e.g., ][]{zsom20121d,gao2018microphysics,loftus2021physics} under a wide variety of different regimes are important for future progress in this area. Finally, and perhaps most importantly, clouds matter to exoplanet observations: most often because they obscure the detection of other spectral features. \cite{komacek2020clouds} and \cite{suissa2020dim} studied the effects of clouds on JWST transmission spectroscopy, and concluded that detection of water vapor will be extremely difficult on water-rich Earth-like planets if spectral features are truncated by a cloud deck. The challenge posed to future observations by clouds is discussed further in Section~\ref{sec:observations}.

\section{CURRENT OBSERVATIONAL CONSTRAINTS}\label{sec:observations}

As discussed in Section~\ref{sec:intro}, observations of rocky planet atmospheres are rapidly developing as new systems are discovered and telescope capabilities grow.  Here we provide a snapshot of the current state of knowledge, and in Section \ref{sec:future}, we discuss prospects for future observations.

{Hubble} and {Spitzer} have been the workhorse observing facilities for rocky planet characterization over the last decade. These space telescopes provide precise, repeatable measurements that are free of contamination from the Earth's atmosphere \citep[e.g.][]{kreidberg2014clouds,ingalls2016}. The {Hubble} Wide Field Camera 3 (WFC3) instrument is capable of low-resolution spectroscopy ($R\sim100$) in the near-infrared between 1.1 and 1.7$\mu$m, and {Spitzer}/IRAC provides photometry in two broad band-passes centered at $3.6$ and $4.5\mu$m \citep{fazio2004spitzer,kimble2008wfc3}. Ground-based telescopes have also been used to measured a handful of transmission spectra at optical wavelengths (see Table~\ref{tab:transmissionspectra}). Direct imaging instrumentation is not yet sensitive to rocky planets, so all observational constraints so far are for transiting systems.

Transmission spectra are currently the most common type of observation. These data are sensitive to molecular and atomic absorption, and also the presence of clouds. For a smaller sample of planets, thermal emission and reflected light observations are also available. To date, rocky planet atmosphere characterization has focused on just 10 of the most observationally accessible systems. Nearly all of these are transiting planets orbiting nearby M-dwarfs. These small stars are advantageous targets because they are (1) more numerous than Sun-like stars, leading to a larger sample; (2) smaller in radius, which increases the signal-to-noise of the planet's atmospheric features; and (3) cooler, leading to Earth-like temperatures closer to the star where planets are statistically more likely to transit. These advantages are colloquially known as the M-dwarf opportunity. A few ultra-short-period planets around larger stars are also accessible (notably 55 Cancri e); we discuss these in Section \ref{subsec:sunlike}. 

\subsection{Planets with M-dwarf host stars}

\begin{figure}[ht]
    \centering
   \includegraphics[width=4in]{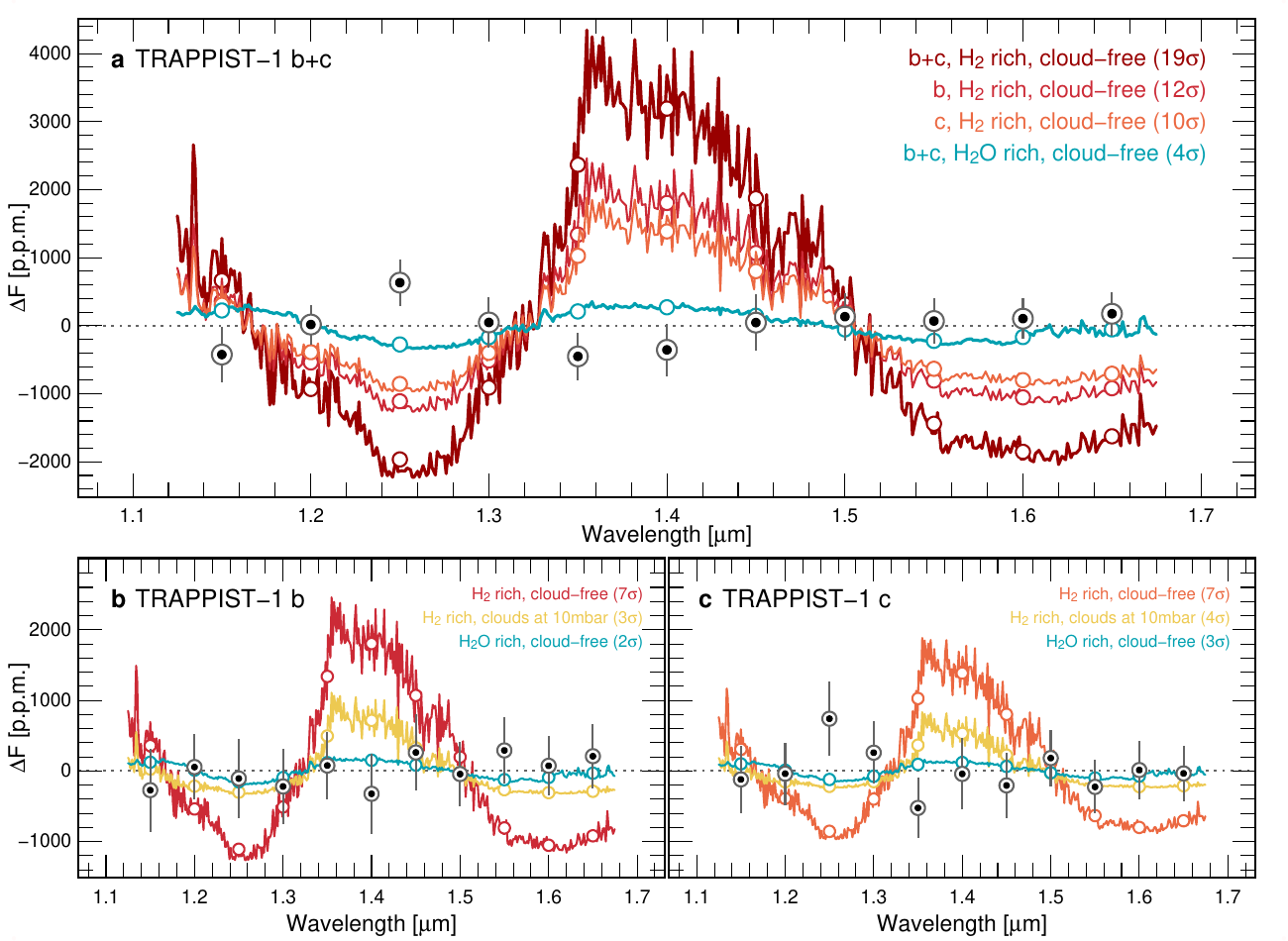}
    \caption{Transmission spectra (black points) for the planets TRAPPIST-1~b and c, measured with the Hubble Space Telescope Wide Field Camera 3 \citep[from ][]{dewit2016}. The top panel shows the combined spectra from both planets, and the bottom panels show each planet spectrum separately. The data are compared to theoretical model predictions for a range of atmospheric compositions (colored lines). A cloud-free, hydrogen-rich atmosphere is ruled out at $7\sigma$ confidence for both planets, but a high mean molecular weight composition of pure \ce{H2O} is marginally consistent with the observations. High altitude clouds (below 10 mbar pressure) or the absence of an atmosphere could also produce a  featureless spectrum.}
    \label{fig:t1bc_spectra}
\end{figure}

Transmission spectra have been measured for nine rocky planets with M-dwarf hosts -- GJ~1132~b, LHS~1140~b, LHS~3844~b, and TRAPPIST-1~b-g.  The measured spectra are generally featureless; i.e., consistent with a constant transit depth over the observed wavelength. Figure\,\ref{fig:t1bc_spectra} shows representative spectra from HST/WFC3 for the two inner planets in the TRAPPIST-1 system \citep{dewit2016}. These featureless spectra are precise enough to rule out cloud-free solar composition (i.e., \ce{H2}-dominated) atmospheres for all but two of the planets studied so far, with a range of significance from 4 to $16\sigma$ (listed in Table~\ref{tab:transmissionspectra}). The spectrum of TRAPPIST-1g has lower signal-to-noise, but still disfavors a solar composition at  $2\sigma$ confidence. In two cases, spectral features in a hydrogen-rich atmosphere were claimed for GJ~1132~b \citep{southworth2017, swain2021}, but both detections were ruled out by subsequent work \citep{diamondlowe2018, mugnai2021, libbyroberts2021}. A Lyman-$\alpha$ transit observation provided further evidence against a hydrogen-rich atmosphere for GJ~1132~b, putting an upper limit on hydrogen outflow rate of $\sim10^9$ g/s  \citep{waalkes2019}.

While cloud-free hydrogen-rich atmospheres are disfavored for the sample of planets studied so far, many other compositions have smaller spectral features that are still consistent with the available data. In particular, high mean molecular weight atmospheres, made primarily of heavier molecules such as \ce{N2}, \ce{H2O} or \ce{CO2}, have spectral features $\sim10$ times smaller than that of a hydrogen-dominated atmosphere. These high mean molecular weight models are well within current measurement uncertainties.   In addition, small spectral features could also arise from high-altitude clouds or hazes that block the transmission of stellar flux through the atmosphere, which have been detected on larger gaseous planets \citep[e.g., ][]{kreidberg2014clouds}. Finally, a featureless spectrum is also expected if there is no atmosphere at all. Higher precision data are needed to distinguish between these possibilities.

Some of the measured spectra show hints of \ce{H2O} features, but these features are likely due to contamination from an imhomogeneous stellar photosphere \citep[expected for cool stars;][]{rackham2018}. For TRAPPIST-1, the combined spectrum of planets b-g shows an inverted water absorption signature that is consistent with \ce{H2O} features in stellar faculae. The transmission spectrum of LHS~1140~b also has a marginally significant ($2.65\sigma$), non-inverted water feature, but this too is consistent with expectations for stellar contamination \citep{edwards21}. For planets around the coolest stars,  \ce{H2O} contamination from an imhomogeneous photosphere may be the norm and should be taken into account in the analysis of the spectra \citep{iyer2020}.

Observation of the planet during secondary eclipse and other phases provides a complementary approach to transmission spectroscopy. These measurements probe the global climate and heat circulation on the planet, which can be used to estimate the surface pressure \citep{selsis2011thermal, koll2019candidates}. Thick atmospheres circulate heat more efficiently, so the higher the surface pressure, the smaller the day-night temperature contrast. Thermal emission and reflection can also constrain the planet's albedo, which can reveal the presence of reflective clouds in an atmosphere and/or surface ices \citep[e.g., ][]{mansfield2019albedo}.

\begin{figure}
    \centering
    \includegraphics[width=14cm]{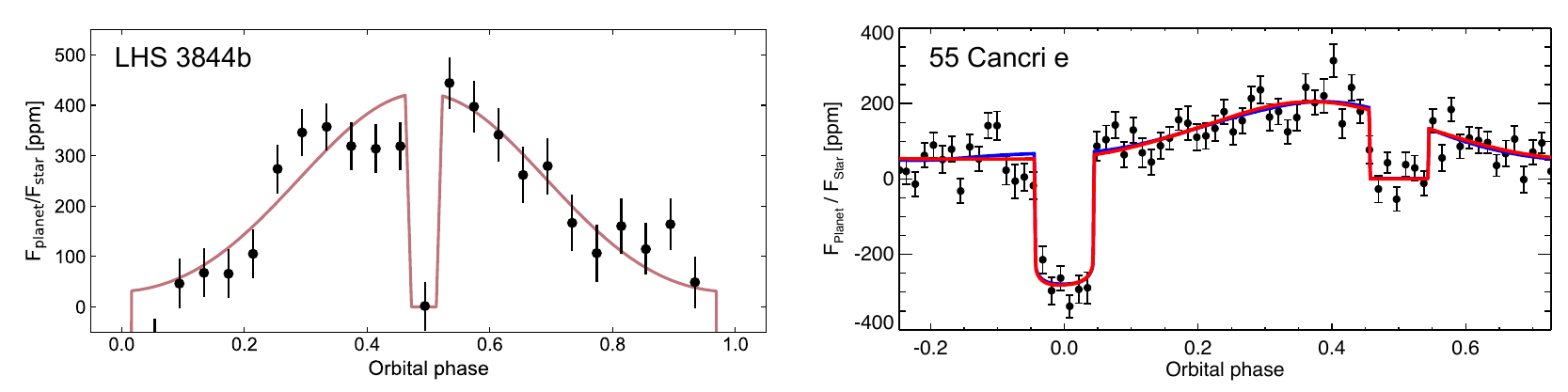}
    \caption{Thermal phase curves and best fit models for the exoplanets LHS~3844~b (left) and 55 Cancri e (right) obtained from Spitzer observations. Adapted from \cite{kreidberg2019absence} and \cite{demory2016map}. The large amplitude, symmetric phase variation of LHS 3844b is consistent with that expected for a bare rock. By contrast, the asymmetric phase curve of 55 Cancri e may be caused by heat circulation in an atmosphere.}
    \label{fig:phase_curve}
\end{figure}

Only one planet with an M-dwarf host has been feasible for thermal emission measurements so far: LHS~3844~b, a 1.3 $R_\oplus$ planet with an 11-hour orbital period. As noted in Section~\ref{sec:atm_dyn}, Spitzer observations at $4.5\mu$m revealed a symmetric, large amplitude thermal phase curve (Figure~\ref{fig:phase_curve}). The large day-night temperature contrast ruled out the presence of a thick ($>10$ bar) atmosphere on the planet \citep{kreidberg2019absence}. Less massive atmospheres are expected to erode by stellar wind (see Section~\ref{sec:atm_escape}) over gigayear timescales, so the planet is most likely a bare rock, with a dark, highly absorptive surface.

\begin{table}[]
\begin{tabular}{llll}
Planet & Result  & Facility & Reference \\
\hline
 GJ 1132 b & solar composition ruled out (16.3$\sigma$) & HST/WFC3 & \cite{libbyroberts2021}  \\
 LHS 1140 b & marginal evidence for H$_2$O ($2.7\sigma$);  & HST/WFC3, & \cite{edwards21},   \\
 \, &          may be due to stellar contamination & Magellan/IMACS \& LDSS3C     & \cite{diamondlowe2020a}             \\
 LHS 3844 b & solar composition ruled out ($5.2\sigma$) & Magellan/LDSS3C &  \cite{diamondlowe2020b}  \\
 TRAPPIST-1 b & solar composition ruled out (7$\sigma$) & HST/WFC3 & \cite{dewit2016}   \\
 TRAPPIST-1 c & solar composition ruled out (7$\sigma$) & HST/WFC3 &  \cite{dewit2016}  \\
 TRAPPIST-1 d & solar composition ruled out (8$\sigma$) & HST/WFC3 &  \cite{dewit2018}  \\
 TRAPPIST-1 e & solar composition ruled out (6$\sigma$) & HST/WFC3  & \cite{dewit2018} \\
 TRAPPIST-1 f &solar composition ruled out (4$\sigma$)  & HST/WFC3 & \cite{dewit2018}   \\
 TRAPPIST-1 g & solar composition disfavored (2$\sigma$) & HST/WFC3 & \cite{dewit2018}
    \end{tabular}\vspace{0.2in}
\caption{Summary of transmission spectroscopy results for rocky planets with M-dwarf host stars.}
\label{tab:transmissionspectra}
\end{table}

\subsection{Ultra-short period planets with Sun-like host stars}\label{subsec:sunlike}

The super-Earth 55 Cancri e is a noteable outlier in the population of small planets that have been studied to date.  It orbits a Sun-like host star ($R_s = 0.96R_{sun}$) every 17.7 hours, resulting in a very high equilibrium temperature (around 2000 K). The planet's bulk composition  is also unusual.  With a radius $R_p = 1.947\pm 0.038\,R_\oplus$, 55 Cancri e is marginally ($1.5\sigma$) larger than predicted for a pure silicate planet with no iron core \citep{crida2018}, implying that it lies on the boundary between rocky and gaseous. Alternatively, it could have an interior enriched in calcium and aluminum, which are less dense than rock  \citep{dorn2019new}. Based on these measurements, it is still uncertain whether the planet has a rocky surface or a thick volatile envelope.
                                                                                
Thanks to its high temperature and very bright host star (visible with the naked eye), 55 Cancri e is a feasible target for atmosphere characterization and has been extensively studied.  The emerging consensus from transit spectroscopy measurements is that 55 Cancri e does not have a low mean molecular weight atmosphere rich in hydrogen and helium.  At the very high temperatures on the planet, any residual hydrogen or helium in the atmosphere is expected to rapidly escape and form an extended exosphere around the planet \citep[as observed for hot gaseous exoplanets; e.g., GJ 436b][]{ehrenreich2015giant}. Transit observations of 55 Cancri e in Lyman $\alpha$ and the metastable helium 1083 nm line have yielded non-detections, putting upper limits on the H/He mass loss rates that are below  those expected from low mean molecular weight compositions \citep{ehrenreich2012, zhang2021helium}. High resolution transmission spectroscopy also put tight upper limits on carbon- and nitrogen-bearing molecules in a hydrogen-rich atmosphere \citep[e.g. $<0.001\%$ HCN by volume;][]{deibert2021}. One exception to these results is the low resolution HST/WFC3 transmission spectrum, which hinted at an HCN feature in a low mean molecular weight atmosphere \citep{tsiaras2016detection}. However, the HCN feature is marginally significant and the retrieved abundance is inconsistent with the upper limit later obtained by \cite{deibert2021}.  Overall, these results agree with theoretical predictions that a hydrogen-dominated atmosphere would be removed by XUV-driven loss on such a highly irradiated planet (see Section~\ref{sec:atm_escape}).
                                             
Thermal emission measurements of 55 Cancri e provide some additional insight into its atmospheric properties. One of the most intriguing observations of the planet is its thermal phase curve, measured by Spitzer at 4.5~\textmu m (Figure~\ref{fig:phase_curve}). The phase curve shows a large day-night temperature contrast and a significantly offset hot spot \citep[$41 \pm 12^\circ$ east of the substellar point;][]{demory2016map}. The maximum hemisphere-averaged temperate is $2700 \pm 270$~K, and the nightside temperature is $1380 \pm 400$~K. These measurements require both poor heat redistribution (to explain the large day-night contrast) and strong dayside circulation (to explain the hot spot offset). This could be due to the atmosphere having a moderate mean molecular weight (\ce{N2} or CO) and a surface pressure of a few bars \citep{angelo2017case, hammond2017linking}. Low viscosity magma flows have also been suggested as a possible explanation for the phase curve, but it is not certain that such flows could transport enough heat to fully explain the data \citep{demory2016map}.

The infrared eclipse depth of 55 Cancri e is also variable over timescales of few months, perhaps due to volcanic activity affecting the temperature structure of the upper atmosphere \citep{demory2016variability}. The nature of this variability and its origins remain poorly constrained, and additional data are needed to resolve this and other open questions surrounding the planet's atmosphere. Fortunately 55 Cancri e is scheduled to be observed by JWST in its first cycle of operations, so additional constraints are forthcoming. 

In addition to 55 Cancri e, there are a few other ultra-short-period rocky planets with secondary eclipse detections. These planets are below 1.6 Earth radii, so are expected to be rocky in composition \citep{rogers2015rocky}. With equilibrium temperatures above 2000 K, their surfaces are expected to be molten, leading to tenuous gas-melt equilibrium atmospheres \citep{schaefer2009chemistry,miguel2011compositions,ito2015}. One well-known example is Kepler-10b, which has a relatively large eclipse depth indicative of a high geometric albedo \citep[$0.61\pm0.17$;][]{batalha2010kepler}. Such a high albedo could be caused by clouds or   unusually reflective lava \citep{rouan2011, essack2020lava}. By contrast, other hot rocky planets appear to have low albedos, consistent with zero \citep{sheets2017albedo}.

\section{FUTURE PROSPECTS}\label{sec:future}

\subsection{The next decade of observations}\label{sec:newobs}

Over the next decade there are numerous new telescopes slated for first light, both in space and on the ground. First up is the James Webb Space Telescope (JWST), scheduled to launch in late 2021. With a 6.5 meter diameter mirror, JWST will have over 7 times the collecting area of Hubble. Its wavelength coverage also extends farther into the infrared, with spectroscopic capability out to 12 microns. In addition to JWST, there are also several next-generation ground-based telescopes, including the  Extremely Large Telescope (ELT), the Giant Magellan Telescope (GMT), and the Thirty Meter Telescope (TMT). These have aperture sizes ranging from 24 to 39 meters, offering an order of magnitude increase in collecting area compared to the current generation of ground-based facilities, and better angular resolution. These new telescopes will enable revolutionary advances in our ability to study the atmospheres of rocky worlds. 

In its first year of operation, JWST is scheduled to observe a sample of roughly 20 rocky planets, over a range of equilibrium temperatures from 170 K (TRAPPIST-1 h) to 2150 K (K2-141b). These observations will provide a first glimpse of the atmospheric properties of rocky worlds in a much greater diversity of environments than we have in the solar system. In particular, JWST observations of transiting planets with M-dwarf hosts will enable the first detection of several strongly absorbing molecules in the near-IR, such as \ce{CO2}, \ce{H2O}, and \ce{CH4}. These molecules may be detectable with a modest number of transits or eclipses (fewer than 10) for a small sample of the most favorable planets \citep{barstow2016trappist,morley2017, lustigyaeger2019trappist, krissansentotton2018,batalha2018jwst, wunderlich2019, gialluca2021jwst}. \ce{H2O} and \ce{CO2} isotopologues may also be detectable for certain atmospheric compositions \citep{lincowski2019observing}. However, species with weaker or narrower absorption features, such as oxygen and ozone, will likely require dozens of transits to detect \citep{lustigyaeger2019trappist}. Small, Earth-like concentrations of these species may be unfeasibly time-intensive to detect.  If clouds are present, they may truncate the amplitude of spectral features and increase the number of transits required by a factor of 10 or more \citep{barstow2016twins,  komacek2020clouds, suissa2020dim}. 

To circumvent the challenge of clouds, an alternative approach is emission spectroscopy. This technique probes deeper in the atmosphere thanks to the direct viewing angle \citep{lustig2019clouds} and is not sensitive to inhomogeneities in the stellar photosphere \citep{rackham2018}. Thermal emission measurements have the added benefit that they are sensitive to atmospheric heat redistribution \citep{selsis2011thermal, kreidberg2016proxima, koll2016temperature} or spectral features from a rocky surface, in cases where no atmosphere is present \citep{hu2012surface}. This technique is most feasible for hot planets, with equilibrium temperatures above 300~K. For these planets, JWST could distinguish between a $\sim1$ bar atmosphere and a bare rock in roughly 100  systems \citep{koll2019candidates}. For a few of the highest signal-to-noise planets, it may also be possible to distinguish between the dominant chemical species in their atmospheres, provided they are optically thick and not too cloudy \citep{morley2017}.

In addition to JWST, there is also a major push for ground-based instrumentation that can characterize rocky planet atmospheres. In the next decade the first 30-m class ground-based telescopes will become available. These facilities will have both higher  spectral resolution and higher angular resolution than any space-based optical/IR telescope. High spectral resolution studies may be capable of detecting a broad range of molecules, including species such as HDO \citep{molliere2019detecting} or even \ce{O2} \citep{snellen2013o2, rodler2014o2}. While \ce{O2} on a temperate planet is a particularly challenging case, expected to require dozens of transit observations even for the most optimistic scenarios, the observing time can be substantially decreased by combining high spectral resolution with high contrast imaging to reduce the stellar light \citep{snellen2015, lovis2017}. Other molecules with a greater number of absorption lines will also be easier to detect, as will features in the atmospheres of hotter planets. High spectral resolution observations will thus play an important complementary role to JWST, particularly if the atmospheres of rocky planets are often cloudy. High resolution observations probe the cores of absorption lines, which form higher in the atmosphere, possibly above the clouds \citep{gandhi2020, hood2020clouds}. Another advantage of high spectral resolution data is that the planet signal is Doppler-shifted relative to stellar absorption lines, which can rule out stellar contamination that can affect low-resolution observations \citep{rackham2018}.  Finally, if the signal-to-noise is high enough, it may even be possible to estimate wind speeds from the Doppler shift of the lines, as has been done for hot Jupiters \citep{snellen2010}. 

\begin{figure}
\includegraphics[width=5.0in]{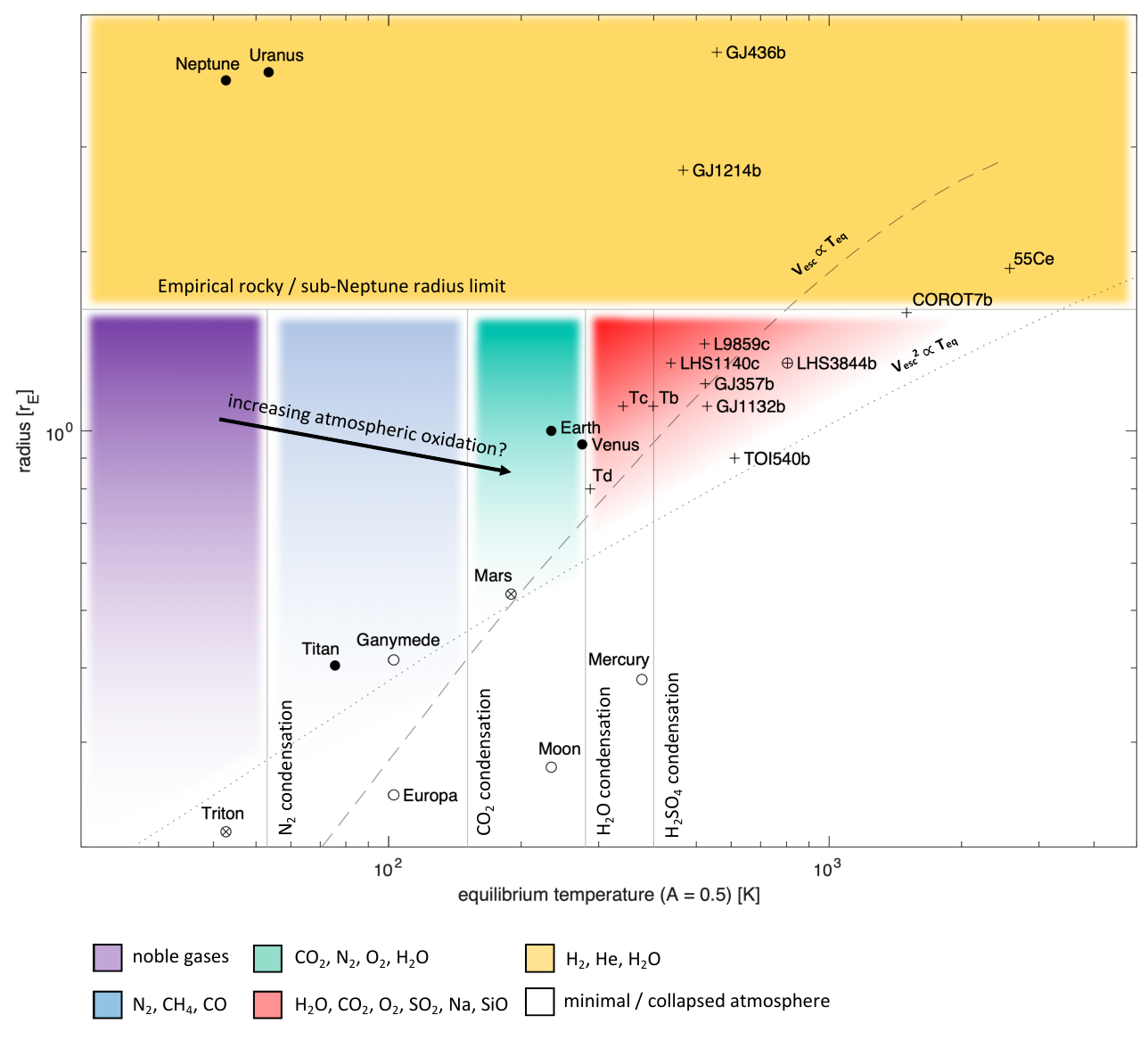}
\caption{Regime diagram of plausible rocky planet atmospheric compositions as a function of equilibrium temperature (assuming an albedo of 0.5) and radius. Solar system planets and moons with stable atmospheres are shown by filled circles, those with collapsed atmospheres are shown by crossed circles, and those without atmospheres are shown by open circles. Selected exoplanets from Table~\ref{tab:best_targets} are shown by the plus symbols, with an open circle for LHS3844b added to indicate its low inferred atmospheric pressure. The legend at the bottom indicates plausible dominant atmospheric constituents for each region of the diagram. The dashed and dotted lines indicate theoretical curves for total atmospheric loss (see main text). Exoplanet name abbreviations are: Tb, Tc and Td for TRAPPIST-1b, c and d, and 55Ce for 55 Cancri e. As can be seen, the rocky exoplanets amenable to near-future characterization occupy a region of parameter space that is not seen in the solar system.} 
\label{fig:phase_diagram}
\end{figure}

\subsection{Testing regime diagrams of rocky planet atmospheres}

The coming wave of new observations will soon make it possible to test regime diagrams of rocky exoplanet atmospheres. One such diagram is shown in Fig.~\ref{fig:phase_diagram}, building on previous work by \cite{forget2014possible} and \cite{zahnle2017cosmic}. 
The diagram axes are equilibrium temperature and planetary radius, on the basis that these two parameters are particularly critical to evolution (although clearly many other parameters are also important). A horizontal line on the plot at $R=1.6 R_\oplus$ divides sub-Neptunes from rocky planets, based on the empirical evidence discussed in Section~\ref{sec:intro} \citep{weiss2014,rogers2015rocky,wolfgang2015}. The vertical lines indicate condensation temperatures of key species\footnote{Condensation temperature is calculated at 300~Pa for \ce{N2}, \ce{CO2} and \ce{H2O}. This choice is somewhat arbitrary, but it ensures that the \ce{H2O} condensation temperature (266~K) is close to the runaway greenhouse equilibrium temperature limit \citep{goldblatt2013low}. Above the condensation temperature of a given species, atmospheric abundance increases rapidly until the surface reservoir is exhausted. The condensation line for \ce{H2SO4} is included because of its importance to atmospheric aerosols, with a value of 400~K used following \cite{lincowski2018evolved}.}, which strongly influences bulk atmospheric composition (Section~\ref{sec:insights}). 

Broadly speaking, planets and satellites in the bottom right of Fig.~\ref{fig:phase_diagram} are expected to be airless. The sloped lines indicate two possible boundaries for this transition. The dashed line is based on the empirical cosmic shoreline of \cite{zahnle2017cosmic}, for which the stellar flux -- escape velocity scaling is given by $F\propto v_{esc}^4 $ (Section~\ref{sec:atm_escape_C}). Given $F \propto T_{eq}^4$, this implies $v_{esc} \propto T_{eq}$. The dotted line follows $v_{esc}^2 \propto T_{eq}$, which would apply for hydrodynamic escape given linear proportionality between $T_{eq}$ and the upper atmosphere temperature, and a transition to rapid escape at some critical $\lambda$ value\footnote{This relationship can be seen by combining the definitions of $\lambda$ (Eqn.~\ref{eq:lambda}) and escape velocity, $v_{esc} \equiv \sqrt{2GM/R}.$}. Both lines assume Earth-like rocky composition to obtain $v_{esc}$ as a function of radius \citep{zeng2019growth} and are normalized to Mars, a planet that has almost but not quite lost its entire atmosphere. 

Perhaps the most striking aspect of Figure~\ref{fig:phase_diagram} is that all the rocky exoplanets shown are in a distinct region of parameter space compared to both solar system bodies and larger exoplanets. Characterizing the atmospheres of these planets will allow us to tackle vital new questions in rocky exoplanet evolution, such as:
\begin{enumerate}
\item What is the dependence of the 1.6~$R_\oplus$ sub-Neptune / rocky boundary on equilibrium temperature? Both exoplanet mass-radius statistics \citep{fulton2018california} and the non-detection of an \ce{H2}-\ce{He} atmosphere on 55 Cancri e suggest highly irradiated planets lose more hydrogen than less irradiated ones, but more data is needed. 
\item Where does the airless boundary for rocky planets occur, and how much scatter in this boundary is driven by other factors such as star type? Existing observations of LHS3844b \citep{kreidberg2019absence} already provide clues, but much more data is needed to characterize this transition comprehensively. 
\item What are the key transitions in atmospheric composition above $T_{eq} = 300$~K? Rocky planets with atmospheres in this regime have no direct analogue in the solar system, and so for the moment we only have theory to guide us. 
\item Is there an observable trend towards more oxidizing atmospheres at higher $T_{eq}$ and lower radius due to H loss, as expected from theory? 
\end{enumerate}
All of these questions provide strong motivation for future observational campaigns.

\subsection{Habitability and biosignatures}

Searching for life on exoplanets is the most exciting future goal of all. The philosophical significance of the discovery of life on another world defies superlatives: it would be the most important breakthrough since the Copernican revolution at least. The challenges to be surmounted remain considerable. Nonetheless, the prospect of characterizing the habitability and potential biospheres of nearby rocky exoplanets is getting ever closer, and enormous strides forward are expected in the coming years. 

In the solar system, we only have a single known example of an inhabited planet, so any attempt to predict habitability for exoplanets is by necessity an extrapolation. While the debate about how to define life is ongoing, most studies of exoplanet habitability use the universal dependence of known life on liquid water and organic carbon chemistry as a starting point. Extensive work has been performed to define the habitable zone, or range of distances from a host star over which a planet may support surface liquid water \citep[e.g., ][]{kasting1993habitable,selsis2007habitable,abe2011habitable,pierrehumbert2011hydrogen,seager2013exoplanet,kopparapu2013habitable,shields2016habitability,ramirez2018more}. The now canonical definition of the habitable zone, which was proposed by \cite{kasting1993habitable}, is the range of distances over which an Earth-like planet can sustain surface liquid water while \ce{CO2} is a non-condensing atmospheric constituent. It therefore roughly corresponds to the region between the \ce{CO2} and \ce{H2O} condensation lines in Fig.~\ref{fig:phase_diagram}. The motivation for this definition is the carbonate-silicate cycle, which is believed to have regulated Earth's surface temperature via \ce{CO2} changes on geologic timescales \citep{walker1981negative}. This definition may require a planet to possess Earth-like plate tectonics \citep{seager2013exoplanet,kasting2014remote}, and \ce{CO2} warming alone cannot explain Mars's evidence for surface liquid water 3--4 billion years ago \citep{wordsworth2016climate}. As a result, many other habitable zone definitions have been proposed using various alternative assumptions (see above references for details). 

Although the validity of the habitable zone concept is still debated \citep[e.g., ][]{moore2017habitable}, it has proved popular in the exoplanet community, and investigations of exoplanet habitability under different scenarios have indirectly helped to improve our understanding of planetary evolution in the solar system. Some recent studies have considered how the habitable zone could be tested observationally.  \cite{bean2017statistical} proposed a statistical approach via low-resolution observations of \ce{CO2} abundances on a large number of rocky exoplanets. However, other authors have suggested the observational challenges will be severe, even for next-generation telescopes \citep{lehmer2020carbonate}. 

An alternative test of rocky exoplanet habitability that avoids \emph{a priori} assumptions about \ce{CO2} regulation is to look for surface liquid water directly. One possibility is to look for signs of ocean glint using a high contrast direct-imaging telescope \citep{robinson2010detecting,lustig2018detecting}. Aqueous chemistry may also be used to constrain surface liquid water abundances indirectly, for example via a planet's sulfur cycle \citep{loftus2019sulfate}, or possibly its ammonia inventory \citep{hu2021unveiling}. While liquid water is essential to Earth-like life, too much water may be detrimental to habitability, because of how it affects volatile sequestration \citep{kitzmann2015unstable,marounina2020internal}, outgassing \citep{krissansen2021waterworlds} and near-surface nutrient availability \citep{wordsworth2013water,glaser2020detectability}. Techniques to determine surface pressure and temperature, and ultimately to detect surface land masses directly, will therefore also be important \citep{cowan2009alien,benneke2012atmospheric}. 

While liquid water is fundamental to habitability, the redox chemistry of a rocky planet is critical to its chances of developing life in the first place. As they lose \ce{H2} to space, many rocky planets may pass through a stage where both liquid water and reducing chemistry allowing formation of prebiotic compounds are present \citep{wordsworth2012transient,luger2015habitable}. However, on planets where \ce{H2O} photolysis and H loss continue once all \ce{H2} is lost, the resulting hyper-oxidation of the atmosphere and surface will cause organic compounds to be rapidly destroyed by oxidants, likely preventing the emergence and survival of Earth-like life. The best targets for life searches may therefore be those planets that have undergone just enough hydrogen escape to remove their primordial \ce{H2} envelopes, but not enough to hyperoxidize their surfaces.

Redox is also critical to the topic of biosignatures \citep{seager2012astrophysical}. Biosignature definition is an important topic that has been covered in detail in several comprehensive recent reviews \citep[e.g., ][]{kaltenegger2017characterize,meadows2018exoplanet,schwieterman2018exoplanet}. In brief, the classic approach to biosignature definition relies on the idea that life pushes planetary atmospheres towards increasing chemical disequilibrium \citep{lovelock1965physical}, leading to the simultaneous presence of oxidizing and reducing atmospheric species (e.g. \ce{O2} and \ce{CH4} on Earth). However, the last few years of research have made it clear that many such gases, particularly \ce{O2}, can also be produced by abiotic processes. Biosignature definition has therefore moved from a focus on close Earth analogues to a broader view, that encompasses the possibility of a wide variety of atmospheric redox states, metabolisms, and signature gases \citep[e.g., ][]{catling2018exoplanet,olson2018atmospheric,sousa2020phosphine}.

Research in this area is ongoing, but two main themes are now clear. First, while no single molecule detection is likely to constitute a `smoking gun' biosignature alone, observations of multiple species in the atmosphere of a single target can be used to build up an increasingly convincing case that a biosphere is present, particularly when additional context is provided. Second, the more we understand how rocky exoplanet atmospheres evolve in the absence of a biosphere, the stronger the case for life will be when biosignatures are detected. The next decade of atmospheric characterization will therefore be a critical phase on the road to detecting life outside our solar system. 

\subsection{Observing biosignatures with a next-generation telescope}

There is broad consensus in the exoplanet community that a next-generation direct imaging mission is needed to detect biosignatures in the atmospheres of Earth-mass planets orbiting Sun-like stars \citep{dalcanton2015, charbonneau2018NAS}. 
 While the next decade will see major steps forward in identifying the basic atmospheric chemistry and climate of rocky exoplanets, almost all of these planets have M-dwarf host stars (Table~\ref{tab:best_targets}). Even for these systems, JWST and the extremely large telescopes (ELTs) will be hard-pressed to identify biosignatures. For example, molecular oxygen may be detectable on a handful of the most accessible planets with intensive observing campaigns (TRAPPIST-1e or Proxima b); however, the simulated detections rely on optimistic assumptions about the oxygen concentration and presence of clouds in the atmosphere. Even if oxygen were detected, it is not a  biosignature on its own. To robustly detect biospheres, a more complete chemical inventory is therefore needed, for a larger sample of planets with a broader range of host star types.

To bridge this gap, several ambitious, next-generation space missions have recently been proposed with the goal of searching for biosignatures on a large sample ($\sim$dozens) of temperate, rocky planets. These include the Habitable Exoplanet Observatory (HABEx) the Large UV/Optical/Infrared Surveyor (LUVOIR), and the Large Interferometer For Exoplanets (LIFE) \citep{gaudi2020,luvoir2019,quanz2021}. The proposed facilities would use a variety of technologies to achieve the high contrast and high angular resolution required for direct imaging of other Earths:  starlight suppression with a coronograph or a starshade, or interferometry. These projects build on the heritage of many previously studied missions, notably the Terrestrial Planet Finder/Darwin concepts \citep{defrere2018}.

Now that the occurrence rate of temperate rocky planets is known \citep{dressing2015, bryson2021}, the time is ripe to pursue a flagship mission to characterize them in more detail. A space-based exoplanet imaging mission was recommended as the top scientific priority by both the US-based 2020 Decadal Survey on Astronomy and Astrophysics \citep{2020Decadal} and the European Space Agency's Voyage 2050 Report.  Specifically, the 2020 Decadal Survey recommended a large ($\sim$6 m aperture) infrared/optical/ultraviolet space telescope to launch in the mid-2040s; such a mission could obtain spectra of 25 potentially habitable exoplanets. Similarly, the Voyage 2050 Report recommended an exoplanet interferometry mission, if it proves to be financially feasible. The 2020 Decadal also recognized the technical challenge and potential high cost of the proposed imaging missions, and laid out a multi-decade path that begins with an aggressive program of technology maturation for the 2020s.

With these recommendations, we are poised to finally address the age-old question: is there alien life around other stars? Exoplanet science has seen enormous progress in the short three decades of its existence, and with an ambitious strategy, hard work from a growing community, and international collaboration, the next 30 years will be even more exciting. We have already learned  that exoplanets are more diverse and varied in their properties than we ever imagined from the small sample in the Solar System. Here we have laid out a snapshot of the current view of rocky exoplanet atmospheres, but if history is any guide, we can expect that this article will be quickly out of date as we continue on the path to characterizing Earth-like worlds. 

\section*{DISCLOSURE STATEMENT}
The authors are not aware of any affiliations, memberships, funding, or financial holdings that might be perceived as affecting the objectivity of this review. 

\section*{ACKNOWLEDGMENTS}
Comprehensively covering such an exciting and fast-moving new field in a single review paper is a great challenge, and we apologize to any colleagues whose work we inadvertently omitted here. We thank Collin Cherubim, Jess Cmiel, Hannah Diamond-Lowe, Feng Ding, Kait Loftus, Laura Schaefer, Jake Lustig-Yaeger, Kevin Zahnle and an anonymous reviewer for discussion and comments that greatly improved this manuscript. This research has made extensive use of NASA's Astrophysics Data System and the NASA Exoplanet Archive. RW acknowledges funding from NSF-CAREER award AST-1847120 and Virtual Planetary Laboratory (VPL) award UWSC10439. LK acknowledges generous support from the Max Planck Society.

\bibliographystyle{apj}
\bibliography{abbr,paper_refs}

\end{document}